\documentclass[12pt]{amsart}
\usepackage{amssymb}
\usepackage{verbatim}
\usepackage[usenames]{color}
\usepackage{hyperref}

\newtheorem{thm}{Theorem}
\newtheorem{prop}[thm]{Proposition}
\newtheorem{la}[thm]{Lemma}

\theoremstyle{definition}
\newtheorem{df}[thm]{Definition}
\newtheorem{conv}[thm]{Convention}

\newtheorem{mdf}{Modification}
\newtheorem{intent}[thm]{Intention}

\theoremstyle{remark}
\newtheorem{ex}[thm]{Example}
\newtheorem{rmk}[thm]{Remark}

\newenvironment{ls}{\begin{itemize}}{\end{itemize}}
\newenvironment{lsnum}{\begin{enumerate}}{\end{enumerate}}
\newenvironment{unn}[1]{\bigskip\noindent\textbf{#1}\quad}{\par\bigskip}
\newenvironment{pf}{\begin{proof}}{\end{proof}}

\newcommand{\scr}[1]{\ensuremath{\mathcal {#1}}}
\newcommand{\bld}[1]{\ensuremath{\mathbf {#1}}}

\newcommand{\ttt}[1]{\ensuremath{\mathtt {#1}}}
\newcommand{\dom}[1]{\ensuremath{{\text{Dom}}(#1)}}
\newcommand{\ran}[1]{\ensuremath{{\text{Range}}(#1)}}

\newcommand{\emp}{\varnothing}
\renewcommand{\phi}{\varphi}

\newcommand{\U}{\ensuremath{\Upsilon}}
\newcommand{\sq}[1]{\ensuremath{\langle#1\rangle}}

\newcommand{\restr}{\mathop{\upharpoonright}}

\newcommand{\notarrow}{\kern .42em\not\kern -.42em\longrightarrow}
\renewcommand{\th}{\ensuremath{{}^{\text{th}}}}
\newcommand{\ite}[3]{\ensuremath{\mathtt{if\ }#1\mathtt{\ then\ }
#2\mathtt{\ else\ }#3\mathtt{\ endif}}}

\newcommand{\DD}{\Delta^+}
\newcommand{\ans}{\dot}

\newcommand{\val}[3]{\ensuremath{\text{Val}(#1,#2,#3)}}
\newcommand{\qval}[3]{\ensuremath{\text{q-Val}(#1,#2,#3)}}
\newcommand{\Issued}{\text{Issued}}
\newcommand{\kand}{\curlywedge}
\newcommand{\kor}{\curlyvee}

\newcommand{\initeq}{\unlhd}

\newcommand{\E}{\text{E}}
\newcommand{\rl}{\ensuremath{\ttt{rl}}}
\newcommand{\ql}[2]{\ensuremath{#1[=:#2]}}
\newenvironment{eatab}
 {\bigskip\noindent\begin{minipage}{\textwidth}\upshape\ttfamily
  \begin{tabbing}mmm\=mmm\=mmm\=mmm\=mmm\=\kill}
 {\end{tabbing}\end{minipage}\bigskip}

\title{Persistent Queries}
\author{Andreas Blass}
\address{Mathematics Department\\
University of Michigan\\
Ann Arbor, MI 48109--1043, U.S.A.}
\email{ablass@umich.edu}
\thanks{Blass is partially supported by NSF grant DMS-0653696 and by
a grant from Microsoft Research}
\author{Yuri Gurevich}
\address{Microsoft Research\\
One Microsoft Way\\
Redmond, WA 98052, U.S.A.}
\email{gurevich@microsoft.com}

\begin{document}

\begin{abstract}
  We propose a syntax and semantics for interactive abstract state
  machines to deal with the following situation.  A query is issued
  during a certain step, but the step ends before any reply is
  received.  Later, a reply arrives, and later yet the algorithm makes
  use of this reply.  By a persistent query, we mean a query for which
  a late reply might be used. Syntactically, our proposal involves
  issuing, along with a persistent query, a location where a late
  reply is to be stored.  Semantically, it involves only a minor
  modification of the existing theory of interactive small-step
  abstract state machines.
\end{abstract}

\maketitle

\section{Introduction}   \label{intro}

An abstract state machine (ASM) describes an algorithm by
telling what it does in any one step.  A run of an ASM is the
result of repeatedly executing the one-step instructions,
possibly interleaved with interventions from the environment.
See \cite{lipari} for details or see Section~\ref{asm} below
for a summary.

Previous theoretical work on ASMs has concentrated on what
happens during a single step.  For example, the papers
\cite{seqth,parthesis, ordinary, general} established, for
various classes of algorithms, the theorem that every algorithm
in the class can be matched, step for step, by an ASM.  Almost
nothing was said there about what happens between steps,
because almost nothing can be said; the environment can make
essentially arbitrary inter-step changes to the state.

Intra-step interaction with the environment, in contrast, was
treated in great detail in \cite{ordinary,general}.  The key
difference from inter-step interaction is that, although the
environment can, during a step, give essentially arbitrary
replies to the algorithm's queries, the effect of these replies
on the state and thus on the future course of the computation
is under the algorithm's control.

In the present paper, we use inter-step interaction to treat an
issue arising out of intra-step interaction, namely the
possibility of a query being answered after the completion of
the step in which the query was issued.  We describe an
extension of ASM syntax to accommodate such late replies, and
we relate it to the ASMs of \cite{general}.

As in \cite{seqth,ordinary,general}, we restrict attention to
small-step --- also known as sequential --- algorithms. The amount of
work that a small-step algorithm performs during any one step is
bounded independently of the state or input. In the rest of the
article, algorithms are by default small-step. In
\cite[Part~I]{ordinary}, we argued that, in principle, the intra-step
interaction of an algorithm with the environment reduces to the
algorithm querying the environment and the environment answering these
queries. In the case of ordinary algorithms \cite{ordinary}, every
query issued during a step needs to be answered before the algorithm
finishes the step. In the general case \cite{general}, however, the
algorithm may finish a step without having all the replies.

If, in such a situation, the reply to a query arrives after the
algorithm's step has ended, then the question arises how to handle the
late reply.  It may happen that the algorithm does not need that late
reply; consider for example an algorithm that issues two queries and
sets $x$ to $1$ when at least one of the two replies arrives. In such
a case, the late reply can simply be ignored or discarded.  But
suppose that the algorithm eventually, at some later step, needs the
late reply. In the framework of \cite{general} where attention is
restricted to one step of an algorithm, the natural solution was
this. If and when the algorithm needs a late reply, it issues an
auxiliary query inquiring whether the reply is in.  Another
possibility, close to current programming practice, is to fork out a
separate computation thread that will wait for the late reply and will
perhaps do some work with the late reply if and when it appears
\cite{futures}. Here we propose a new solution that does not require
additional queries or additional computation threads.

We base our discussion on the model of interactive computation
introduced and analyzed in \cite{general}, which we review in
Section~\ref{asm}.  This model differs from the earlier, more
special model of \cite{ordinary} in two ways, one of which is
the possibility of completing a computation step without
waiting for replies to all the queries issued during the
step.\footnote{The other is that the algorithm can take into
account the order in which replies are received.}  It is this
possibility that opens the door to the topic of late replies
and their subsequent use by the algorithm.  The primary purpose
of this paper is to describe an ASM model that incorporates
such \emph{persistent queries} and their \emph{late replies}.

The paper is organized as follows.  We begin in
Section~\ref{problem} with some examples showing the relevance
of late replies.  In Section~\ref{soln}, we briefly describe
our proposed extension of the traditional ASM syntax to handle
persistent queries and late replies.  This description is
intended to convey the general idea, without presupposing
details about the traditional syntax and semantics.  Those
details and the associated semantical notions are reviewed in
Section~\ref{asm}, in preparation for a more careful
presentation of our proposal.  Section~\ref{impat} discusses in
more detail the possibility of finishing a step while some of
that step's queries remain unanswered. Finally,
Section~\ref{locate} presents in detail our ASM model for this
situation, and shows how these ASMs can be represented in the
model from \cite{general}.

As indicated above, we limit ourselves here to small-step algorithms,
i.e., algorithms that work in discrete steps (as opposed to
distributed algorithms where there may be no clear notion of (global)
step because agents act asynchronously) and do only a bounded amount
of work per step, with the bound depending only on the algorithm, not
on the input or state (as opposed to, for example, massively parallel
algorithms where the number of available processors may be increased
according to the input size).

There are several justifications for this limitation.  First,
many of the algorithms used in practice are small-step.

Second, even in massively parallel or distributed algorithms,
the individual processors or agents are usually small-step
algorithms.  Communication between agents is, from the point of
view of any one agent, an interaction with its environment.  An
important motivation for developing a general model for
interaction between an algorithm and its environment is this
situation where the algorithm under consideration is one agent
--- a small-step algorithm --- while the environment includes the
other agents.

Third, small-step algorithms are the only class of algorithms
for which the general theory of interactive algorithms has been
worked out in detail and for which interactive ASMs have been
proved to be adequate to capture all algorithms of the class
\cite{ordinary,general}. For parallel algorithms, the analogous
work has been done only in the absence of intra-step
interaction \cite{parthesis}, and the case of distributed
algorithms remains entirely in the domain of future work. Thus,
the foundation on which we shall build in the present paper is
currently available only for small-step algorithms.

Finally, it is reasonable to expect that what we do here for
small-step algorithms will suggest how to do the analogous
tasks for broader classes of algorithms, once the necessary
framework is in place.
What we do here may also be useful in extending our work on parallel
algorithms \cite{parthesis} to include external interactions, because
it allows greater flexibility in handling the flood (or trickle) of
replies that a parallel algorithm might receive all at once.

\specialsection*{PART I: AN IMPROVED INTERACTIVE SMALL-STEP ASM MODEL}

In this part we explain a new model of interactive small-step abstract
state machines that allows us to handle persistent queries. The
explanation covers the syntax and its intended meaning. In the second
part of the paper we cover the formal semantics of the new model in
full detail.

\section{Persistent Queries and Late Replies}   \label{problem}

As indicated earlier, we are concerned in this paper with
providing an ASM formalism that conveniently handles the
following situation: An algorithm issues a query during a
certain step, but finishes the step without getting an answer
to that query. The answer arrives later and is then used in
some subsequent step of the algorithm.

By a \emph{late reply}, we mean a reply from the environment to
a query $q$, reaching the algorithm after the completion of the
step in which $q$ was issued.  If a late reply to $q$ can
influence the subsequent work of the algorithm, then we call
$q$ a \emph{persistent query}.

\begin{rmk}
The most natural meaning of ``can influence'' in the preceding
sentence involves what can actually happen in runs of the algorithm.
Like other run-time properties, persistence is then undecidable in
general.  That undecidabilty does no harm to our work in this paper.
On the other hand, when writing programs, one is faced with the need
to decide which queries should be considered persistent and treated by
the methods of this paper.  For this purpose, one should interpret
``can influence'' to mean that the programmer does not know with
certainty that a late answer will never be used.  It does no harm if a
program treats a query as persistent even when, at run time, it turns
out not to be persistent.
\end{rmk}

\begin{ex}   \label{broker1}
We revisit Example~3.3 of \cite[Part~I]{general}.  In that
example, a broker has a block of shares to sell and offers the
entire block to two clients.  As soon as he gets a positive
reply from either client, he sells all the shares to that
client. (The situation where positive replies from both clients
reach the broker simultaneously is discussed in \cite[Part~I,
Example~3.20]{general}, but it need not concern us here.)
Suppose that the broker has sold the shares to client~A and
completed his step (with an update to his state, recording the
sale) without having received any reply from client~B.  Later,
he gets a reply from B, who also wants to buy the shares.  He
should then tell B, ``Sorry, I already sold the shares to
someone else, whose acceptance of my offer reached me before
yours.''  Thus, the actions of the broker (regarded as an
algorithm) take into account B's reply, even though the reply
came after the completion of the step in which the associated
query (the offer to sell the shares) was issued.  So that query
is persistent.
\end{ex}

\begin{ex}  \label{pollster1}
A pollster sends questionnaires to many people.  Being a
small-step algorithm, the pollster sends the questionnaires a
few at a time, so the sending occupies numerous steps.  Later,
the filled-in questionnaires arrive  and the pollster processes
them. Usually, a questionnaire will be filled in and returned
only after the end of the (pollster's) step in which it was
sent out. So the filled-in questionnaires are late replies, and
the associated queries, the original, blank questionnaires, are
persistent queries.

Even if one of the respondents is so quick that the pollster
gets the reply in the same step in which he issued the query
(so we are dealing with a traditional reply, as in
\cite{ordinary,general}, not a late reply), the pollster will
probably want to postpone processing this reply until after he
finishes mailing all the questionnaires.  More generally, an
algorithm may well treat all replies the same, whether they are
late or not.
\end{ex}

How should persistent queries and late replies be treated in
the context of ASMs?

Recall (from \cite{lipari} or \cite{ordinary} or \cite{general}
--- see Section~\ref{asm} below for a review) how queries arise
and how their answers are used in the computation done by an
ASM.  Queries are produced by terms $f(t_1,\dots,t_n)$ in the
ASM program, where $f$ is an external function symbol.  The
queries and replies from any single step of the computation
form a history (in \cite{general}, an answer function in
\cite{ordinary}), which is empty at the beginning of a step and
gradually grows as queries are issued and answered.  The
history influences the algorithm's actions (issuing additional
queries, ending the step, updating the state) during that step,
but it is reset to empty for the start of the next step.  This
is in accordance with the general principle --- intuitively the
definition of ``state'' --- that the state must include all the
information from past steps that can influence the future
progress of the computation.

As in previous work \cite{ordinary, general}, we adopt the so-called
Lipari convention, namely that if the same external function symbol
occurs several times in an ASM program, and if its arguments at
different occurrences happen to evaluate to the same elements in a
particular state, then all those occurrences result in only a single
query during a single step of the algorithm.  For a discussion of
alternative conventions and of our reasons for adopting the Lipari
convention, see \cite[Part~II, Section~4]{ordinary}.

\begin{rmk}
  One of those alternative conventions, the must-vary convention, is
  commonly used in practice.  This convention requires all occurrences
  of external functions in a program to produce different queries,
  even if they involve the same function with the same arguments.  The
  idea is that, whenever a query is issued, an additional component,
  an ID, is added automatically, and all these IDs are distinct.
  Thus, even if two queries look the same to the ASM, the IDs make
  them distinct.  (In \cite{ordinary}, the must-vary convention
  applied to the queries issued within a single step, but we naturally
  take it to also apply to queries from different steps.  All the IDs
  are distinct, whether from the same step or not.)  It appears that,
  given a suitable formalization of ASM semantics under the must-vary
  convention, what we do in this paper would work under that
  convention as well.  We do not attempt to develop the must-vary
  version of the theory here, but we shall add occasional remarks
  about how this convention would affect our discussion.
\end{rmk}

With this rough description of
the situation (see Section~\ref{asm} for a detailed
description), we turn to the question of handling persistent
queries and late replies in the context of ASMs.

Perhaps the first approach that comes to mind is that, when an
algorithm wants to use a late reply to a previously issued
query $q$, it simply re-issues $q$.  Then the late reply to the
old $q$ would appear, in the history of the later step, as the
reply to the new $q$.

\begin{rmk}
  Under the must-vary convention, this approach would not arise, since
  there would be no such thing as re-issuing a query.
\end{rmk}

The trouble with this approach is that re-issuing an old query
already has a different meaning: It is an entirely new query,
not related (in general) to the previous query.  In particular,
if $q$ is issued and answered at some step and then issued
again at a later step, it may get an entirely different reply
the second time.  In other words, the histories that occur in
different steps of a computation are not required to agree in
any way. Recall that each step of an algorithm's computation
begins with an empty history and gradually builds up to a
larger history as queries are issued and answered, but at the
end of the step, its history disappears, so there is no
connection between the query $q$ issued at two different steps.
Formally, this fact is incorporated in the the definition of
coherence and the Step Postulate in \cite{general} (reviewed in
Section~\ref{asm} below), which make no allowance for any
influence of the histories of earlier steps. Informally, the
same fact is a consequence of the general principle  that all
the information from the computation's past that can affect its
future must be in the state, not in some other memory of
histories from previous steps.

An algorithm, having issued $q$ in some earlier step but having
received no answer, might well want to both use a late reply to
that $q$ and also issue $q$ anew for a possibly different
reply.  Obviously, this situation cannot be modeled by using a
re-issued $q$ to represent looking for a late reply.

The same difficulty can also be seen in the two examples above.
If the broker looked for a late reply from client B by
re-issuing the query, then this would look to B like a new
offer to sell a (possibly different) block of shares.
Similarly, for a pollster to look at the replies he has
received is quite different from sending out the questionnaires
again.

A second approach also uses queries whereby the algorithm looks for
late replies, but these queries will not be repetitions of the
original queries.  Instead, this approach is similar to the use of
implicit queries, which was introduced in \cite[Part~I,
Section~2]{ordinary} as a way to represent an algorithm's paying
attention to unsolicited information from the environment.  (See also
\cite[Part~II, Example~5.14]{ordinary} and \cite[Part~I,
Remark~3.7]{general}.)  The idea here is that, when it wants to use a
late reply to a query $q$, the algorithm is, in effect, asking the
environment to provide that late reply, if one exists.  That is, the
algorithm issues a query asking, ``What late reply, if any, has been
received for the query $q$?''

\begin{rmk}
  Unlike the first approach, the second makes sense in the must-vary
  context.  The following paragraph would, however, be modified.
  Instead of needing tags in case the same query is issued at several
  previous steps (an impossibility under must-vary), the algorithm
  would need to know the IDs that were attached to its previous
  queries.
\end{rmk}

Actually, this query needs to be more detailed.  The query $q$
could have been issued at several earlier steps, and these
occurrences of $q$ would be treated by the environment as
distinct queries,
which could receive different answers.  So
late replies might be available for several of these
occurrences.  The algorithm needs to say which occurrence it
wants.  So the implicit query might have the form ``What late
reply, if any, has been received for the query $q$ that I
issued in step $n$?''  To avoid the need for both the algorithm
and the environment to count steps, the algorithm might assign
some tags to persistent queries at the time it issues them, and
inform the environment about the tags, so that it can later ask
``What late reply, if any, has been received for the query $q$
that I issued with tag $t$?''

A version of this approach was suggested in \cite[Part~I,
Section~2]{ordinary}, even though the algorithms of that paper
never finished a step with unanswered queries.  Nevertheless,
the possibility of an query-reply pair spanning several steps
was addressed as follows.  The query should be regarded as a
simple message to the environment, whose answer received in the
same step is an uninformative ``OK,'' and the ``real'' answer
in some later step should be regarded as a message from the
environment, which is formally regarded as the reply to an
implicit query ``I'm willing to receive a message.''  This
version leaves it up to the environment to say which old query
it is answering with its new message.

The use of new queries to request late replies to old queries
has some drawbacks. It requires additional work from the
environment, namely storing all late replies until the
algorithm asks for them, and then delivering them immediately.
This produces a mismatch between the ASM model and what would
ordinarily happen in practice.  A real environment would
probably deliver a late reply as soon as it is available and
expect the algorithm to deal with it from then on.

If a reply is not yet available when requested, the algorithm
might well keep issuing the same request, step after step, and
just reading all those requests would be a burden for the
environment.  This hardly matters as long as we take the
algorithm's point of view and regard the environment as given.
But it would matter in a distributed algorithm, where an
agent's environment consists of other agents and the algorithms
executed by those agents would have to include ways of handling
a barrage of queries for which the answer isn't available.

Such approaches also clash, on a more philosophical level, with
the standard ASM notion of state.  Once a late reply is
available, it is something that resulted from the past steps of
the computation and may be relevant to the future; so it ought
to be part of the state.

A third approach, quite common in practice, is the use of futures, new
threads created to allow the algorithm (or the parent thread) to
proceed without waiting for late replies to persistent queries. A
future could receive a late reply and either report it to the parent
thread (or some other thread) or do some other work with it.  It could
also do additional work both before and after the late reply arrives.
This approach would take us out of the realm of small-step algorithms
for two reasons.  First, the new threads need not be synchronized with
the parent thread or with each other, so we would no longer have a
global state advancing step by step.  Second, even if we demanded
synchronization, a situation could arise where an algorithm has issued
a great many queries at various steps in the past (only a few queries
at any one step, if the algorithm is small-step) and has a great many
futures waiting for the replies.  If these futures do any computing
while they wait, the total work they do might not be bounded.  Since
we want to remain in the framework of small-step algorithms, we do not
adopt futures as our method of handling late replies.

We therefore prefer a fourth approach, in which a late reply is
recorded directly in the algorithm's state.  The
next section explores this approach in somewhat more detail.

\section{ASMs With Persistent Queries}      \label{soln}

In this section, we discuss the last approach mentioned above for
handling persistent queries and late replies.  When a late reply
becomes available, the environment should record it in the state of
the algorithm.

The environment's action of recording the late reply, since it
takes place without a new query from the algorithm, is an
inter-step interaction.  It directly updates the algorithm's
state, without any action by the algorithm.
So this update cannot occur earlier than first inter-step moment after
the reply becomes available.  It might occur later, if the environment
is busy with other tasks or if communication is slow.  Fortunately,
this makes no difference difference to our discussion (though it may
make a difference to the efficiency of the algorithm).  Indeed, from
the point of view of an algorithm (or ASM) it makes no difference if a
late reply, received at the start of a certain step, was actually
available much earlier to the environment; the algorithm simply
doesn't see such availability.  As far as the algorithm is concerned,
the only notion of ``available'' is ``delivererd to me by the
environment.''

If this method of communication between the environment and the
algorithm is to succeed, they must agree as to where, in the
algorithm's state, a late reply to a particular query is to be
recorded.  The environment must know where to put the reply,
and the algorithm must know where to find the reply when
needed.  We propose that this agreement be achieved as follows.
Whenever it issues a query for which a late reply might be
relevant (a persistent query), the algorithm should give the
environment, along with the query, a \emph{reply location},
where any late reply to this query should be recorded.

As in \cite{lipari} (and all subsequent work on ASMs), we take
\emph{location} to mean a pair \sq{f,\bld a} where $f$ is a
function symbol of the algorithm's vocabulary and $\bld a$ is a
tuple of elements of the state, an $n$-tuple if $f$ is $n$-ary.

Recall (from \cite{ordinary,general} or see Section~\ref{asm} below)
that a query is a tuple of elements of the disjoint union
$X\sqcup\Lambda$, where $X$ is (the underlying set of) the state and
$\Lambda$ is a set of labels.  If the function symbol $f$ is among the
labels, then a location \sq{f,\bld a} is almost a query.  ``Almost''
because the location is \sq{f,\sq{a_1,\dots,a_n}} while the query is
\sq{f,a_1,\dots,a_n}; we shall ignore such bracketing distinctions in
the future and write as if locations are queries.

It is not enough, however, for the algorithm to issue, along
with any persistent query, its reply location as a second query
(an \ttt{Output} in the sense of \cite{ordinary} or an
\ttt{issue} in the sense of \cite{general}, to which the
environment gives an automatic, immediate, and uninformative
reply).  The algorithm must tell the environment which reply
location goes with which query.  After all, the algorithm might
issue many queries simultaneously.

The simplest way for the algorithm to convey the necessary
information to the environment is to issue, along with any
persistent query $q$, a second query that contains both $q$ and
the reply location.  We adopt, by convention, the following
format for this second query.  It is the concatenation of three
sequences:
\begin{ls}
\item the query $q$,
\item the one-term sequence \sq{\rl}, and
\item the reply location  $l$.
\end{ls}
Here the special label \rl\ (abbreviating ``reply location'')
marks where the query ends and the reply location begins (and
indicates that there \emph{is} a reply location, i.e., that
this is not just another query); we assume that this marker
\rl\ is chosen to be distinct from all other labels used by the
algorithm.

Here a simplification is possible, if the environment is
willing to cooperate.  The information in the original query
$q$ is repeated in the first part of the additional message
\sq{q,\rl,l} that specifies the reply location $l$.  So there
is no real need to issue $q$; it would suffice to issue
\sq{q,\rl,l} if the environment is smart enough to interpret it
as follows: Regard the part before \rl\ as a query in the
traditional sense, but, if the reply is late, then put it into
the location given after \rl.

As a further simplification, we adopt the convention that the
reply to a persistent query should be put into its reply
location even if the reply arrives during the step in which the
query was issued.  We impose no requirement, however, on how soon such
a reply is put into the reply location.  It need not happen at the end
of the step in which the query was issued; it might happen at the end
of some later step.  The reason for this flexibility is that we do not
wish to impose requirements on how fast the environment works, or even
on the relative speed of different parts of the environment.  Thus,
one part of the environment may be able to provide an immediate reply
directly to the algorithm (as in \cite{ordinary}) while the part of
the environment responsible for inter-step changes to the algorithm's
state is slower.  Fortunately, this flexibility does no harm to our
theory.

Instead of having the environment put on-time replies into the reply location, we could program algorithms so that, when a persistent query is issued and answered during the same step, the algorithm writes the reply into the reply location.  Our convention relieves the algorithm of this duty, assigning it to the environment instead.

Does this reassignment unduly burden the environment?  One can
argue that it actually makes the environment's job easier.  If
only late replies are to be written to the reply location, then
the environment must watch the step-by-step progress of the
algorithm's work, in order to know whether a particular reply
is late.  With our convention, the environment need not monitor
the algorithm in such detail; all replies to persistent queries
go into the assigned reply locations.  The only difference
between on-time and late replies is that the former are seen by
the algorithm, in its history (or answer function), without
having to wait until the end of the step.

We propose the following ASM syntax for generating the combined
queries --- persistent query combined with reply location. (The
same syntax could also be used in a framework where $q$ and
\sq{q,\rl,l} are issued separately.)  Suppose the query $q$
results from a term $g(\bld u)$ in an ASM.  So $g$ is an
$m$-ary external function symbol for some $m$ and \bld u is an
$m$-tuple of terms $u_i$; $q$ results from inserting the values
(in the algorithm's current state) of the $u_i$'s in the
template associated to $g$.
(Recall from \cite[Part~II, Section~4.2]{ordinary} that a template is
like a query but with placeholders instead of elements of the state.
An ASM provides, for each external function symbol $g$, a template
$\hat g$.  The query issued by $g$ with arguments $u_i$ is obtained by
replacing the placeholders in $\hat g$ by the values of the $u_i$'s.
See also Section~4 below.)  Suppose further that the desired location
for late replies is \sq{f,\bld a}. The components $a_j$ of the tuple
\bld a must be the values, in the current state, of some terms $t_j$,
in order for the algorithm to be able to refer to them.  Then the
algorithm can specify the desired location by means of the term
$f(\bld t)$. To say, in an ASM program, that the algorithm should ask
the query arising from $g(\bld u)$ and to specify \sq{f,\bld a} as its
reply location, we write
\[
\ql{g(\bld u)}{f(\bld t)}.
\]
For human readability, the brackets indicate that the main
query here is produced by $g(\bld u)$, and the
reverse-assignment notation $=:$ indicates that $f(\bld t)$ is
to be read as specifying a location (like the left side of an
update rule written with $:=$) and that the value to be put
there is the (eventual) value of $g(\bld u)$.

  In the situation described here, since \sq{f,\bld a} is a location,
  the function symbol $f$ must be in the state vocabulary, not an
  external function symbol.  In fact, we require all function symbols
  in the terms \bld t to be from the state vocabulary also.  That is,
  the ASM should not need to issue queries in order to determine reply
  locations.  This requirement arises from the combination of two
  circumstances.  First, a reply location for a query should be
  determined when the query is issued, not in some later step.  (It
  would be a serious problem if the query location were determined
  only after the arrival of the reply that should go into this
  location.)  So any queries arising from external function symbols in
  \bld t need to be answered in the current step, not later.  Second,
  it turns out that whether a query must be answered in the current
  step depends only on the context in which it appears in the ASM
  program.  (The relevant contexts are timing guards, guards built
  with Kleene connectives, and issue rules.  In all other contexts,
  queries must be answered in the current step.  This will be proved
  formally in Proposition~\ref{source} below.) But a persistent query
  $g(\bld u)$ and its reply location $f(\bld t)$ share the same
  context.  So if the latter must be answered in the current step, so
  must the former.  And then the former doesn't need a reply location.

\begin{rmk}
We could relax this requirement and allow \bld t to issue queries
provided we have some assurance, from a source other than the context
in the ASM program, that these queries will be answered in the current
step.  Such assurance could come from knowledge about the
environment.  It could also come from other parts of the ASM program.
A simple example of the latter possibility is given by the program

\begin{eatab}
  if $t=t$ then\\
\>if $c\prec \ql{g(u)}{f(t)}$ then\\
\>\>$x:=0$\\
\>endif\\
endif.
\end{eatab}

\noindent Here $t$, $c$, and $g$ are external but $f$ is in the state
vocabulary.  A step of this algorithm can finish without a value for
$g(u)$, provided $c$ has a value.  But it cannot finish without a
value for $t$, because of the guard $t=t$ (whose sole purpose is to
require that $t$ have a value).
 \end{rmk}

\begin{ex} \label{broker2} Consider a simplified version of the broker
  example as detailed in \cite[Part~I, Example~3.20]{general}; the
  purpose of the simplification is to avoid hiding the currently
  relevant topic, persistent queries and late replies, in a sea of
  other considerations.  We regard the broker's offers to the two
  clients (whom we name 0 and 1) as given by nullary external
  functions $q_0$ and $q_1$ (instead of ternary functions having the
  stock, the number of shares, and the price as arguments), and we
  assume the broker breaks ties (when he gets positive answers from
  both clients simultaneously) in favor of client 0 (rather than
  non-deterministically or randomly).  Also, we assume that, as long
  as the broker has received no answer from either client, or has a
  negative answer from one client and no answer from the other, he
  simply waits.  The resulting algorithm is represented by the
  following ASM, in the notation of \cite[Part~II]{general}. We assume
  that the dynamic function symbols $s_0$ and $s_1$ are used to
  indicate a sale to client 0 or 1, respectively, so they have the
  value \ttt{false} in initial states.

\begin{eatab}
if $\neg$ Halt then\\
\>do in parallel\\
\>\>if $s_0=s_1=$\,false\,$\kand\,
q_0=\ttt{true}\,\kand\,(q_0\preceq
q_1\kor q_1=\ttt{false})$\\
\>\>\>then $s_0:=$\,true endif\\
\>\>if $s_0=s_1=$\,false\,$\kand\,
q_1=\ttt{true}\,\kand\,(q_1\prec
q_0\kor q_0=\ttt{false})$\\
\>\>\>then $s_1:=$\,true endif\\
\>\>if $s_0=s_1=$\,false\,\,$\kand\,q_0=\ttt{false}\,\kand\,q_1=\ttt{false}$\\
\>\>\>then skip endif\\
\>\>Halt := true\\
\>enddo\\
endif
\end{eatab}

The so-called Kleene conjunction $\kand$ and Kleene disjunction
$\kor$ that are used in this ASM program are like ordinary conjunction
$\land$ and disjunction $\lor$ except that $p\kand q$ is false as soon
as one conjunct is false, even if the other is undefined, and dually
for $\kor$.  For more details, see \cite[Part~II,
Section~2.3]{general} or Section~\ref{asm} below.

\begin{conv}  \label{halt}
  In future examples, we shall omit ``$\ttt{if}\ \neg\ttt{Halt\
    then}$'' and the assocated ``\ttt{endif}'', adopting instead the
  convention that an ASM program is to be executed repeatedly until
  \ttt{Halt} becomes true.  This convention supersedes the one from
  \cite{lipari} that the iteration continues until there is no change
  of state from one step to the next.  The new convention allows an
  algorithm to continue waiting for a late reply without making any
  changes to its state.
\end{conv}

Now suppose, as in Example~\ref{broker1} above, we want the
algorithm to respond to a late reply from a losing client with
a letter explaining that the shares have already been sold.  We
assume (again for simplicity, to avoid hiding the relevant
issues) that the broker's vocabulary contains nullary symbols
$l_0$ and $l_1$ denoting appropriate letters to the two
clients.  And we assume that it also has nullary symbols $a_0$
and $a_1$, initially denoting \ttt{undef}, to be used as the
reply locations.  Then the modified algorithm, which behaves
like the one above but also sends the appropriate letter, is
given in our proposed syntax by the following ASM.

\begin{eatab}
do in parallel\\
\>if $s_0=s_1=$\,false\,$\kand\,
q_0=\ttt{true}\,\kand\,(q_0\preceq\ql{q_1}{a_1}\kor
q_1=\ttt{false})$\\
\>\>then $s_0:=$\,true endif\\
\>if $s_0=s_1=$\,false\,$\kand\,
q_1=\ttt{true}\,\kand\,(q_1\prec\ql{q_0}{a_0}\kor
q_0=\ttt{false})$\\
\>\>then $s_1:=$\,true endif\\
\>if $s_0=s_1=$\,false\,$\kand\,q_0=\ttt{false}\,\kand\,q_1=\ttt{false}$\\
\>\>then skip endif\\
\>if $s_0=\ttt{true}\land a_1=$\,true then issue$(l_1)$ endif\\
\>if $s_1=\ttt{true}\land a_0=$\,true then issue$(l_0)$ endif\\
\>if $(a_0=\ttt{true}\lor a_0=\ttt{false})\land(a_1=\ttt{true}\lor
a_1=\ttt{false})$\\
\>\>then Halt $:=$ true endif\\
 enddo
\end{eatab}

The first two lines have been modified by attaching reply locations
$a_i$ to the two query-producing terms $q_i$.  (It doesn't really
matter which occurrence of $q_i$ is annotated with $a_i$. We chose to
use the occurrence that is primarily responsible for the possibility
of finishing the step without a reply.)  Two new lines have been
added, containing instructions for issuing the appropriate letter to
the losing client.  The last line makes the algorithm end its run when
both clients have answered; until then, even if the shares have been
sold to one client, it waits for an answer from the other client.
\end{ex}

This example serves to illustrate a general feature of our
notation.  The part of the program that tells what to do with
late replies to the queries $q_i$ does not mention those
queries at all.  Rather, it mentions the locations $a_i$ where
the late replies are to be found.  The executor of the
algorithm need not remember, when using a late reply, the query
that it answers; only the location of the late reply is
relevant, and it is used like any other location in the state.

\begin{rmk}
The example also has a somewhat special property, namely that it
doesn't need modes.  It is common, in ASM programs, to use certain
nullary, dynamic symbols as modes, to keep track of what sort of work
the algorithm is currently doing.  In the present example, there would
be two modes, one indicating that the broker is waiting for a positive
reply in order to sell the stock, and one indicating that the stock
has been sold to one of the clients but the broker may still need to
send a letter to the other client.  It is often convenient to include
such modes and update them explicitly in an ASM program.  In the
present case, however, this would be redundant, as the first mode
is already described by $s_0=s_1=\ttt{false}$ and the second mode by
the negation of this.
\end{rmk}

\begin{ex}  \label{pollster2}
Consider the pollster example, \ref{pollster1}.  Let us assume
that the pollster sends out $N$ questionnaires, numbered from 0
to $N-1$, that the replies will be numbers, and that the
desired output is the sum of all these numbers.  For
simplicity, we also assume that the questionnaires are sent one
at a time and that all the replies eventually arrive, though
perhaps late and out of order; our pollster algorithm will keep
running without producing an output until all the replies have
been received and added.  The pollster first sends out all the
questionnaires (using an internal variable $i$ to keep track of
where he is in this process) and then goes through all the
replies, adding them one at a time (re-using $i$ to keep track
of this process as well).  We describe what the pollster does
as an ASM, using the following vocabulary.  As already
indicated, $i$ is a dynamic, nullary symbol ranging from 0 to
$N-1$ and indexing the queries and their replies; it is
initially 0. An additional dynamic, nullary symbol
\ttt{all}-\ttt{sent}, initially \ttt{false}, tells whether all the
questionnaires have been sent.  Unary functions $q$ and $l$
send each $i$ to the $i\th$ questionnaire $q(i)$ and its reply
location $l(i)$.  The initial value of $l(i)$ is \ttt{undef}
for each $i$.   A dynamic, nullary function \ttt{sum},
initially 0, represents, at each step, the sum of the replies
that have been added so far. Elementary arithmetic is assumed
to be available, particularly $+$, $<$, and names for specific
numbers. Here is the ASM:

\begin{eatab}
do in parallel\\
\>if all-sent = false $\land\, i<N$ then do in parallel\\
\>\>issue(\ql{q(i)}{l(i)})\\
\>\>$i:=i+1$\\
\>enddo endif\\
\>if all-sent = false $\land\, i=N$ then do in parallel\\
\>\>$i:=0$\\
\>\> all-sent $:=$ true\\
\>enddo endif\\
\>if all-sent = true $\land\, i<N\land\, l(i)\neq\ttt{undef}$
then
do in parallel\\
\>\>sum $:=$ sum $+\,l(i)$\\
\>\>$i:=i+1$\\
\>enddo endif\\
\>if all-sent = true $\land\, i=N$ then Halt $:=$ true endif\\
enddo
\end{eatab}

\noindent Recall here Convention~\ref{halt} that a run of the ASM ends
when \ttt{Halt} becomes true; until then the program is executed
repeatedly.  We also assume that \ttt{Halt} is initially false, so
that the program runs.
\end{ex}

\begin{rmk}
When justifying the ``query and reply'' paradigm for intra-step
interaction in \cite[Part~I, Section~2]{ordinary}, we wrote
that, if an algorithm sends a message to the outside world
without expecting a reply, then this situation can be modeled
by imagining an automatic, immediate, and uninformative reply
``OK,'' essentially just an acknowledgment that the message was
sent.  The \ttt{Output} rules in \cite[Part~II]{ordinary} and
the \ttt{issue} rules in \cite[Part~II]{general} were
introduced to produce such messages.   There is, however,
nothing in the official semantics in \cite{ordinary} or
\cite{general} to require the environment to produce only
``OK'' as a reply to such queries.  Although an \ttt{issue}
rule cannot make use of any nontrivial information provided by
its reply, nothing prohibits the existence of such information.

In fact, there are situations where such nontrivial information
is to be expected, for example in

\begin{eatab}
do in parallel\\
\>x := q\\
\>issue(q)\\
enddo
\end{eatab}

\noindent (where \ttt{q} is an external nullary symbol and
\ttt{x} an internal dynamic one).  In this (admittedly silly)
program, the query produced by the \ttt{issue} line is also
produced, with the intention of using its reply, by the update
rule \ttt{x\ :=\ q}.

Following the official semantics given for ASMs in
\cite[Part~II]{general}, we make no special assumptions about
the replies to queries that result from \ttt{issue} rules.
These replies can be any elements of the state, just as for any
other queries.
\end{rmk}

This convention was used in Example~\ref{pollster2}, because
the queries produced by \ttt{issue(\ql{q(i)}{l(i)})} are the
questionnaires, whose replies should be the numbers stored in
locations $l(i)$ and then added.

This example also used the earlier convention, whereby replies
go into the reply locations even if they are not late.  Without
this convention, the ASM program would have to include
instructions whereby, if an answer to $q(i)$ appears in the
same step in which the query was issued, the algorithm would
put that answer into location $l(i)$.

\begin{rmk}
  Futures can provide a particular way of implementing our approach to
  late replies.  A future that simply waits for a late reply and, when
  one arrives, writes it into the appropriate reply location thereby
  accomplishes what we require of the environment.  Nevertheless,
  there is a conceptual difference.  By assigning to the environment
  the task of putting the late replies into the proper locations, we
  maintain sequentiality of the algorithm.  By assigning the same task
  to futures, a part of the algorithm, one enters the more complex
  domain of asynchronous, distributed algorithms.
\end{rmk}

\specialsection*{PART II: THE DETAILS}

\section{Interactive Abstract State Machines}  \label{asm}

We now begin a more formal treatment of ASMs with persistent
queries.  We build on the ASM model described in
\cite{general}.  In the present section, we summarize the
material from \cite{general} that we need here. This summary
also serves to explain things that were taken for granted in
the preceding sections.  We do not, however, repeat the
extensive discussion offered in \cite{general} to motivate and
explain the model.

We begin by recalling the definitions, conventions, and
postulates for interactive small-step algorithms.  This
material is taken from \cite[Part~I, Section~3]{general}.

\begin{unn}{States Postulate:} The algorithm determines
\begin{ls}
\item a finite vocabulary $\Upsilon$,
\item a nonempty set \scr S of \emph{states}, which are
    $\Upsilon$-structures,
\item a nonempty subset $\scr I\subseteq\scr S$ of
    \emph{initial
  states},
\item a finite set $\Lambda$ of \emph{labels} (to be used in
    forming
  queries).
\end{ls}
\end{unn}

As in earlier papers, we use the following conventions
concerning vocabularies and structures.

\begin{conv}   \label{vocab}
\mbox{}
  \begin{ls}
    \item A vocabulary $\Upsilon$ consists of function symbols with
    specified arities.
    \item Some of the symbols in $\Upsilon$ may be marked as
    \emph{static}, and some may be marked as \emph{relational}.
    Symbols not marked as static are called \emph{dynamic}.
    \item Among the symbols in $\Upsilon$ are the logic names: nullary
    symbols \ttt{true}, \ttt{false}, and \ttt{undef}; unary
    \ttt{Boole}; binary equality; and the usual propositional
    connectives.  All of these are static and all but \ttt{undef} are
    relational.
  \item An $\Upsilon$-structure $X$ consists of a nonempty base set,
  usually denoted by the same symbol $X$,
  and interpretations of all the function symbols $f$ of $\Upsilon$ as
  functions $f_X$ on that base set.
  \item In any $\Upsilon$-structure, the interpretations of \ttt{true},
    \ttt{false}, and \ttt{undef} are distinct.
    \item In any $\Upsilon$-structure $X$, the interpretations of
    relational symbols are functions whose values lie in
    $\{\ttt{true}_X,\ttt{false}_X\}$.
  \item In any $\Upsilon$-structure $X$, the interpretation of
    \ttt{Boole} maps $\ttt{true}_X$ and $\ttt{false}_X$ to
    $\ttt{true}_X$ and everything else to $\ttt{false}_X$.
    \item In any $\Upsilon$-structure $X$, the interpretation of
    equality maps pairs of equal elements to $\ttt{true}_X$ and all
    other pairs to $\ttt{false}_X$.
    \item In any $\Upsilon$-structure $X$, the propositional connectives
    are interpreted in the usual way when their arguments are in
    $\{\ttt{true}_X,\ttt{false}_X\}$, and they take the value
    $\ttt{false}_X$ whenever any argument is not in
    $\{\ttt{true}_X,\ttt{false}_X\}$.
\item We may omit subscripts $X$, for example from \ttt{true}
    and \ttt{false}, when there is no danger of confusion.\qed
  \end{ls}
\end{conv}

\begin{df}
A \emph{potential query} in state $X$ is a finite tuple of
elements of $X\sqcup\Lambda$.  A \emph{potential reply} in $X$
is an element of $X$. \qed\end{df}

Here $X\sqcup\Lambda$ means the disjoint union of $X$ and
$\Lambda$. So if they are not disjoint, then they are to be
replaced by disjoint isomorphic copies.  We shall usually not
mention these isomorphisms; that is, we write as though $X$ and
$\Lambda$ were disjoint.

\begin{df}       \label{ans-fn}
An \emph{answer function} for a state $X$ is a partial map from
potential queries to potential replies.  A \emph{history} for
$X$ is a pair $\xi=\sq{\ans\xi,\leq_\xi}$ consisting of an
answer function $\ans\xi$ together with a linear pre-order
$\leq_\xi$ of its domain. By the \emph{domain} of a history
$\xi$, we mean the domain \dom{\ans\xi} of its answer function
component, which is also the field of its pre-order component.
\qed\end{df}

Recall that a \emph{pre-order} of a set $D$ is a reflexive,
transitive, binary relation on $D$, and that it is said to be
\emph{linear} if, for all $x,y\in D$, $x\leq y$ or $y\leq x$.
The equivalence relation defined by a pre-order is given by
$$
x\equiv y\iff x\leq y\leq x.
$$
The equivalence classes are partially ordered by
$$
[x]\leq[y]\iff x\leq y,
$$
and this partial order is linear if and only if the pre-order
was.

We also write $x<y$ to mean $x\leq y$ and $y\not\leq x$.
(Because a pre-order need not be antisymmetric, $x<y$ is in
general a stronger statement than the conjunction of $x\leq y$
and $x\neq y$.)  When, as in the definition above, a pre-order
is written as $\leq_\xi$, we write the corresponding
equivalence relation and strict order as $\equiv_\xi$ and
$<_\xi$.  The same applies to other subscripts and
superscripts.

We use histories to express the information received by the
algorithm from its environment during a step.  The answer
function part $\dot\xi$ of a history tells what replies the
environment has given to the algorithm's queries, and the
pre-order part $\leq_\xi$ tells in what order these replies
were received.  Specifically, if $q$ is in the domain of $\xi$,
then $\dot\xi(q)$ is the environment's answer to the query $q$.
If $p,q\in\dom{\xi}$ and $p<_\xi q$, this means that the answer
${\ans\xi}(p)$ to $p$ was received strictly before the answer
${\ans\xi}(q)$ to $q$.  If $p\equiv_\xi q$, this means that the
two answers were received simultaneously.

We emphasize that the timing we are concerned with here is
logical time, not physical time.  That is, it is measured by
the progress of the computation, not by an external clock.  In
particular, we regard a query as being issued by the algorithm
as soon as the information causing that query (in the sense of
the Interaction Postulate below) is available.  This is why we
include, in histories, only the relative ordering of replies.
The ordering of queries relative to replies or relative to each
other is then determined.  The logical time of a query is the
same as the logical time of the last of the replies needed to
cause that query.

\begin{df}
Let $\leq$ be a pre-order of a set $D$.  An \emph{initial
segment} of $D$ with respect to $\leq$ is a subset $S$ of $D$
such that whenever $x\leq y$ and $y\in S$ then $x\in S$.  An
\emph{initial segment} of $\leq$ is the restriction of $\leq$
to an initial segment of $D$ with respect to $\leq$.  An
\emph{initial segment} of a history \sq{{\ans\xi},\leq_\xi} is
a history \sq{{\ans\xi}\restr S,\leq_\xi\restr S}, where $S$ is
an initial segment of \dom{\ans\xi}\ with respect to
$\leq_\xi$.  (We use the standard notation $\restr$ for the
restriction of a function or a relation to a set.)  We write
$\eta\initeq\xi$ to mean that the history $\eta$ is an initial
segment of the history $\xi$.  If $q\in D$, then we define two
  associated initial segments as follows.
\begin{align*}
(\leq q)&=\{d\in D:d\leq q\}\\
(< q)&=\{d\in D:d< q\}. \qed
\end{align*}
\end{df}

\begin{unn}{Interaction Postulate}
For each state $X$, the algorithm determines a binary relation
  $\vdash_X$, called the \emph{causality relation}, between finite
  histories and potential queries.
\end{unn}

The intended meaning of $\xi\vdash_Xq$ is that, if the algorithm's
current state is $X$ and the history of its interaction so far (as
seen by the algorithm during the current step) is $\xi$, then it will
issue the query $q$ unless it has already done so in the current step.
When we say that the history so far is $\xi$, we mean not only that
the environment has given the replies indicated in $\ans\xi$ in the
order given by $\leq_\xi$, but also that no other queries have been
answered.  Thus, although $\xi$ explicitly contains only positive
information about the replies received so far, it also implicitly
contains the negative information that there have been no other
replies.  Of course, if additional replies are received later, so that
the new history has $\xi$ as a proper initial segment, then $q$ is
still among the issued queries, because it was issued at the earlier
time when the history was only $\xi$.  This observation is formalized
as follows.

\begin{df}  \label{iss-pend}
 For any state $X$ and history $\xi$, we define sets of queries
\begin{align*}
\Issued_X(\xi)&=\{q:(\exists\eta\initeq\xi)\,\eta\vdash_Xq\}\\
\text{Pending}_X(\xi)&=\Issued_X(\xi)-\dom{\ans\xi}.\qed
\end{align*}
\end{df}

Thus, $\Issued_X(\xi)$ is the set of queries that have been
issued by the algorithm, in state $X$, by the time the history
is $\xi$, and $\text{Pending}_X(\xi)$ is the subset of those
that have, as yet, no replies.

The following definition describes the histories that are
consistent with the given causality relation.  Informally,
these are the histories where every query in the domain has a
legitimate reason, under the causality relation, for being
there.

\begin{df}
  A history $\xi$ is \emph{coherent}, with respect to a state
  $X$ or its associated causality relation $\vdash_X$, if
  \dom{\ans\xi} is finite and
  \[
  (\forall q\in\dom{\ans\xi})\,q\in\Issued_X(\xi\restr(<q))
\]
\qed\end{df}

\begin{rmk}
In \cite[Part~I, Definition~3.12]{general}, the definition of
coherence did not require \dom{\ans\xi} to be finite; instead,
it had the weaker requirement that the linear order of
$\equiv_\xi$-classes induced by $\leq_\xi$ is a well-order. The
stronger requirement of finiteness was, however, deduced later
from the Bounded Work Postulate for all attainable histories;
see \cite[Part~I, Corollary~3.28]{general}.  Since we omit such
deductions here, it seems clearer to build finiteness
explicitly into the notion of coherent history.
\end{rmk}

\begin{df}
A history $\xi$ for a state $X$ is \emph{complete} if
$\text{Pending}_X(\xi)=\emp$. \qed\end{df}

The terminology reflects the fact that, if a complete history has
arisen in the course of a computation, then there will be no further
interaction with the environment during this step.  No further
interaction can originate with the environment, because no queries
remain to be answered.  No further interaction can originate with the
algorithm, since $\xi$ and its initial segments don't cause any
further queries.  So the algorithm must either terminate its run
(successfully) if \ttt{Halt} becomes true, or proceed to the next step
(by updating its state), or fail.
The next definitions and postulates describe
these end-of-step matters.
They do not explicitly mention termination
(other than by failure), but this is covered anyway, since updates are
covered and termination amounts to an update of \ttt{Halt} to the value
\ttt {true}.

\begin{df}
A \emph{location} in a state $X$ is a pair \sq{f,\bld a} where
$f$ is a dynamic function symbol from $\Upsilon$ and $\bld a$
is a tuple of elements of $X$, of the right length to serve as
an argument for the function $f_X$ interpreting the symbol $f$
in the state $X$.  The \emph{value} of this location in $X$ is
$f_X(\bld a)$.  An \emph{update} for $X$ is a pair $(l,b)$
consisting of a location $l$ and an element $b$ of $X$.  An
update $(l,b)$ is \emph{trivial} (in $X$) if $b$ is the value
of $l$ in $X$.  We often omit parentheses and brackets, writing
locations as \sq{f,a_1,\dots,a_n} instead of
\sq{f,\sq{a_1,\dots,a_n}} and writing updates as \sq{f,\bld
a,b} or \sq{f,a_1,\dots,a_n,b} instead of $(\sq{f,\bld a},b)$
or $(\sq{f,\sq{a_1,\dots,a_n}},b)$. \qed\end{df}

The intended meaning of an update \sq{f,\bld a,b} is that the
interpretation of $f$ is to be changed (if necessary, i.e., if
the update is not trivial) so that its value at $\bld a$ is
$b$.

\begin{unn}{Step Postulate --- Part A}
The algorithm determines, for each state $X$, a set $\scr F_X$
of \emph{final histories}.  Every complete, coherent history
has an initial segment (possibly the whole history) in $\scr
F_X$.
\end{unn}

Intuitively, a history is final for $X$ if, whenever it arises
in the course of a computation in $X$, the algorithm completes
its step, either by failing or by executing its updates and
proceeding to the next step or terminating the run if \ttt{Halt} has
become true.

\begin{df}   \label{att-def}
  A history for a state $X$ is \emph{attainable} (in $X$) if it
  is coherent and no proper initial segment of it is final.
\qed\end{df}

The attainable histories are those that can occur under the
given causality relation and the given choice of final
histories.  That is, not only are the queries answered in an
order consistent with $\vdash_X$ (coherence), but the history
does not continue beyond where $\scr F_X$ says it should stop.

\begin{unn}{Step Postulate --- Part B}
  For each state $X$, the algorithm determines that certain histories
  \emph{succeed} and others \emph{fail}.  Every final, attainable
  history either succeeds or fails but not both.
\end{unn}

\begin{df}
  We write $\scr F_X^+$ for the set of successful final histories and
  $\scr F_X^-$ for the set of failing final histories.
\end{df}

The intended meaning of ``succeed'' and ``fail'' is that a
successful final history is one in which the algorithm finishes
its step and performs a set of updates of its state, while a
failing final history is one in which the algorithm cannot
continue --- the step ends, but there is no next state, not
even a repetition of the current state. Such a situation can
arise if the algorithm computes inconsistent updates.  It can
also arise if the environment gives inappropriate answers to
some queries.

\begin{unn}{Step Postulate --- Part C}
For each attainable history $\xi\in\scr F^+_X$ for a state $X$,
the algorithm determines an \emph{update set} $\DD(X,\xi)$,
whose elements are updates for $X$.  It also produces a
\emph{next state} $\tau(X,\xi)$, which
\begin{ls}
  \item has the same base set as $X$,
  \item has $f_{\tau(X,\xi)}(\bld a)=b$ if $\sq{f,\bld
  a,b}\in\DD(X,\xi)$, and
  \item otherwise interprets function symbols as in $X$.
\end{ls}
\end{unn}

\begin{conv}
  In notations like $\scr F_X$, $\scr F^+_X$, $\scr F^-_X$,
  $\DD(X,\xi)$, and $\tau(X,\xi)$, we may omit $X$ if only one $X$ is
  under discussion.  We may also add the algorithm $A$ as a
  superscript if several algorithms are under discussion.
  \qed\end{conv}

Any isomorphism $i:X\cong Y$ between states can be extended in
an obvious, canonical way to act on queries, answer functions,
histories, locations, updates, etc.  We use the same symbol $i$
for all these extensions.

\begin{unn}{Isomorphism Postulate}
Suppose $X$ is a state and $i:X\cong Y$ is an isomorphism of
$\Upsilon$-structures.  Then:
\begin{ls}
  \item $Y$ is a state, initial if $X$ is.
  \item $i$ preserves causality, that is, if $\xi\vdash_Xq$ then
  $i(\xi)\vdash_Yi(q)$.
  \item $i$ preserves finality, success, and failure, that is, $i(\scr
  F_X^+)=\scr F_Y^+$ and $i(\scr F_X^-)=\scr F_Y^-$.
  \item $i$ preserves updates, that is, $i(\DD(X,\xi))=\DD(Y,i(\xi))$
  for all histories $\xi$ for $X$.
\end{ls}
\end{unn}

\begin{conv}
  In the last part of this postulate, and throughout this paper, we
  adopt the convention that an equation between possibly undefined
  expressions is to be understood as implying that if either side is
  defined then so is the other.
\qed\end{conv}

\begin{unn}{Bounded Work Postulate}
\mbox{}
  \begin{ls}
    \item There is a bound, depending only on the algorithm, for the
    lengths of the tuples in $\Issued_X(\xi)$ , for all states $X$ and
    final, attainable histories $\xi$.
    \item There is a bound, depending only on the algorithm, for the
    cardinality $|\Issued_X(\xi)|$, for all states $X$ and final,
    attainable histories $\xi$.
    \item There is a finite set $W$ of $\Upsilon$-terms (possibly
    involving variables), depending only on the algorithm, with the
    following property.  Suppose $X$ and $X'$ are two states and $\xi$
    is a history for both of them.  Suppose further that each term in
    $W$ has the same value in $X$ as in $X'$ when the variables are
    given the same values in \ran{\ans\xi}.  Then:
    \begin{ls}
      \item If $\xi\vdash_Xq$ then $\xi\vdash_{X'}q$ (so in particular
      $q$ is a query for $X'$).
      \item If $\xi$ is in $\scr F_X^+$ or $\scr F_X^-$, then it is
      also in $\scr F_{X'}^+$ or $\scr F_{X'}^-$, respectively.
      \item $\DD(X,\xi)=\DD(X',\xi)$.
    \end{ls}
  \end{ls}
\end{unn}

\begin{df}   \label{alg-def}
  An \emph{interactive, small-step algorithm} is any entity satisfying
  the States, Interaction, Step, Isomorphism, and Bounded Work
  Postulates.
\qed\end{df}

Since these are the only algorithms under consideration in most
of this paper, we often omit ``interactive, small-step.''

\begin{df}
  A set $W$ with the property required in the third part of the
Bounded Work Postulate is called a \emph{bounded exploration
witness} for the algorithm.  Two pairs $(X,\xi)$ and
$(X',\xi)$, consisting of states $X$ and $X'$ and a single
$\xi$ that is a history for both, are said to \emph{agree} on
$W$ if, as in the postulate, each term in $W$ has the same
value in $X$ as in $X'$ when the variables are given the same
values in \ran{\ans\xi}. \qed\end{df}

This completes our review of the notion of interactive
small-step algorithm, as defined in \cite[Part~I]{general}.
This notion will be slightly modified in Section~\ref{locate}
to accommodate our proposal for handling persistent queries. A
modification is needed because, when an algorithm issues a
combination \sq{q,\rl,l} of a query $q$ and a reply location
$l$, the reply (if received in the same step) is a reply to
$q$.  So it is $q$, not \sq{q,\rl,l}, that should appear in the
domain of the history.  See Section~\ref{locate} for more
details.

We now turn to the notion of abstract state machine (ASM) from
Part~II of \cite{general}.  ASMs describe algorithms, and the
main result of \cite{general} is that all algorithms (as
defined above) are behaviorally equivalent (in a very strong
sense defined in \cite[Part~I, Section~4]{general}) to ASMs. We
begin our review of ASMs by summarizing the syntactic
definitions from \cite[Part~II, Section~2]{general}; afterward,
we shall also summarize the semantics.

An ASM uses a vocabulary $\Upsilon$, subject to
Convention~\ref{vocab}, and a set $\Lambda$ of labels as in our
discussion of algorithms above. In addition, it has an
\emph{external vocabulary} E, consisting of finitely many
\emph{external function symbols}.  These symbols are used
syntactically exactly like static, non-relational symbols from
$\U$, but their semantics will be quite different.  If $f$ is
an $n$-ary external function symbol and $\bld a$ is an
$n$-tuple of arguments from a state $X$, then the value of $f$
at $\bld a$ is not stored as part of the structure of the state
but is obtained from the environment as the reply to a query.
If the history contains no reply to this query, then $f$ has no
value at \bld a.

\begin{df}
  The set of \emph{terms} is the smallest set containing
  $f(t_1,\dots,t_n)$ whenever it contains $t_1,\dots,t_n$ and $f$ is
  an $n$-ary function symbol from $\U\cup\E$.  (The basis of this
  recursive definition is, of course, given by the 0-ary function
  symbols.)  A \emph{Boolean term} is a term of the form $f(\bld t)$
  where $f$ is a relational symbol.  \qed\end{df}

\begin{conv}
  By $\Upsilon$-\emph{terms}, we mean terms built using the function
  symbols in $\Upsilon$ and variables.  These are terms in the usual
  sense of first-order logic for the vocabulary $\Upsilon$.  They occur,
  for example, in the Bounded Work Postulate as elements of the bounded
  exploration witness.   Terms as
  defined above, using function symbols from $\Upsilon\cup\E$ but not
  using variables, will be called \emph{ASM-terms} when we wish to
  emphasize the distinction from $\Upsilon$-terms.  A term of the form
  $f(\bld t)$ where $f \in \E$ is called a \emph{query-term}.
  \end{conv}

We introduce timing explicitly into the formalism with the
notation $(s\preceq t)$, which is intended to mean that the
replies needed to evaluate the term $s$ arrived no later than
those needed to evaluate $t$.  As explained in \cite{general},
$\preceq$ differs from function symbols in that $s\preceq t$
can have a truth value even when only one of $s$ and $t$ has a
value.

We also use a version of the Boolean connectives with similar
behavior, so that, for example, a disjunction counts as true as
soon as one of the disjuncts is, even if the other disjunct has
no truth value.  This behavior characterizes the connectives of
Kleene's strong three-valued logic.   We use the notations
$\kand$ and $\kor$ for the conjunction and disjunction of this
logic; the traditional conjunction and disjunction, which have
values only when both constituents do, will continue to be
written $\land$ and $\lor$.

\begin{df}
  The set of \emph{guards} is defined by the following recursion.
  \begin{ls}
    \item Every Boolean term is a guard.
    \item If $s$ and $t$ are terms, then $(s\preceq t)$ is a guard.
    \item If $\phi$ and $\psi$ are guards, then so are
    $(\phi\kand\psi)$, $(\phi\kor\psi)$, and $\neg\phi$.
  \end{ls}
\qed\end{df}

\begin{df}
The set of ASM \emph{rules} is defined by the following
recursion.
\begin{ls}
\item If $f\in\U$ is a dynamic $n$-ary function symbol, if
  $t_1,\dots,t_n$ are terms, and if $t_0$ is a term that is Boolean if
  $f$ is relational, then
$$
f(t_1,\dots,t_n):=t_0
$$
is a rule, called an \emph{update} rule.
\item If $f\in \E$ is an external $n$-ary function symbol and
    if
  $t_1,\dots,t_n$ are terms, then
$$
\ttt{issue}\ f(t_1,\dots,t_n)
$$
is a rule, called an \emph{issue} rule.
\item \ttt{fail} is a rule.
\item If $\phi$ is a guard and if $R_0$ and $R_1$ are rules,
    then
$$
\ttt{if\ } \phi\
\ttt{\ then\ } R_0 \ttt{\ else\ } R_1 \ttt{\ endif}
$$
is a rule, called a \emph{conditional} rule.  $R_0$ and $R_1$
are its \emph{true} and \emph{false branches}, respectively.
\item If $k$ is a natural number (possibly zero) and if
  $R_1,\dots,R_k$ are rules then
$$
\ttt{do\ in\ parallel\ }R_1,\dots,R_k\ttt{\ enddo}
$$
is a rule, called a \emph{parallel combination} or \emph{block}
with the subrules $R_i$ as its \emph{components}.
\end{ls}
\qed\end{df}

We may omit the end-markers \ttt{endif} and \ttt{enddo} when
they are not needed, for example in very short rules or in
programs formatted so that indentation makes the grouping
clear.

The correspondence between external function calls and queries
is mediated by a template assignment, defined as follows.

\begin{df}
For a fixed label set $\Lambda$, a \emph{template} for $n$-ary
  function symbols is any tuple in which certain positions are filled
  with labels from $\Lambda$ while the rest are filled with the
  \emph{placeholders} $\#1,\dots,\#n$, occurring once
  each.  We assume that these placeholders are distinct from all
  the other symbols under discussion ($\U\cup \E \cup\Lambda$).
  If $Q$ is a template for $n$-ary functions, then we write
  $Q[a_1,\dots,a_n]$ for the result of replacing each placeholder
  $\#i$ in $Q$ by the corresponding $a_i$.
  \qed\end{df}

Thus if the $a_i$ are elements of a state $X$ then
$Q[a_1,\dots,a_n]$ is a potential query in $X$.

\begin{df}
For a fixed label set and external vocabulary, a \emph{template
assignment} is a function assigning to each $n$-ary external
function symbol $f$ a template $\hat f$ for $n$-ary functions.
\qed\end{df}

The intention, which will be formalized in the semantic
definitions below, is that when an ASM evaluates a term
$f(t_1,\dots,t_n)$ where $f\in \E$, it first computes the
values $a_i$ of the terms $t_i$, then issues the query $\hat
f[a_1,\dots,a_n]$, and finally uses the answer to this query as
the value of $f(t_1,\dots,t_n)$.

By assigning templates to external function symbols, rather than to
their occurrences in a rule, we incorporate into our framework the
``Lipari convention'' of \cite[Part~II, Section~4.3]{ordinary}.  This
means that, if an external function symbol has several occurrences in
an ASM program and if its arguments have the same values at these
occurrences, then only a single query will be issued in any one step
as a result of all of these occurrences.
See Sections~4.3--4.6 of
\cite[Part~II]{ordinary} for a discussion of alternative conventions,
and see \cite[Part~III, Section~7]{ordinary} for additional
information comparing these conventions.

\begin{df}   \label{asm-prog-def}
  An \emph{interactive, small-step, ASM program} $\Pi$ consists of
  \begin{ls}
    \item a finite vocabulary $\U$,
    \item a finite set $\Lambda$ of labels,
    \item a finite external vocabulary $\E$,
    \item a rule $R$, using the vocabularies $\U$ and
    $\E$, the \emph{underlying rule} of $\Pi$,
    \item a template assignment with respect to $\E$ and $\Lambda$.
  \end{ls}
\end{df}

\begin{conv}
  We use the following abbreviations:
\begin{align*}
    &(s \prec t)   &&\text{for} &&\neg(t \preceq s),  \\
    &(s \approx t)   &&\text{for} &&(s \preceq t) \kand (t \preceq s),\\
    &(s \succeq t) &&\text{for} &&(t \preceq s), \text{and}\\
    &(s \succ t)   &&\text{for} &&(t \prec s)
\end{align*}
We abbreviate the empty block \ttt{do\ in\ parallel\ enddo} as
\ttt{skip}. We may omit parentheses when no confusion results.
\qed\end{conv}

This completes the syntax of ASMs; we turn next to the
semantics, as presented in \cite[Part~II, Section~3]{general}.
We treat terms, guards, and rules in turn.  Their semantics are
defined in the presence of a state $X$, a template assignment,
and a history $\xi$.

The semantics of terms specifies, by induction on terms $t$,
the queries that are caused by $\xi$ under the associated
causality relation $\vdash^t_X$ and sometimes also a value
$\val tX\xi\in X$.  In the case of query-terms, the semantics
may specify also a query called the query-value $\qval tX\xi$.
Evaluation of a query-term $t$ should first issue the query
$\qval tX\xi$; the reply, if any, to this query is the actual
value $\val tX\xi$ of $t$.

\begin{df}[Semantics of Terms]   \label{term-val-def}
  Let $t$ be the term $f(t_1,\dots,t_n)$.
  \begin{lsnum}
   \item If \val{t_i}X\xi\ is undefined for at least one $i$, then
    \val tX\xi\ is also undefined, and $\xi\vdash^t_Xq$ if and only
    if $\xi\vdash^{t_i}_Xq$ for at least one $i$. If $f \in
    \E$ then $\qval tX\xi$ is also undefined.

   \item If, for each $i$, $\val{t_i}X\xi=a_i$ and if $f\in\U$,
    then $\val tX\xi=f_X(a_1,\dots,a_n)$, and no query $q$ is
    caused by $\xi$.

   \item If, for each $i$, $\val{t_i}X\xi=a_i$, and if $f\in \E$, then
    $\qval tX\xi$ is the query $\hat f[a_1,\dots,a_n]$.

   \begin{ls}

    \item If $\qval tX\xi = q \in \dom{\ans \xi}$, then $\val
    tX\xi=\ans\xi(q)$, and no query is caused by $\xi$.

    \item If $\qval tX\xi = q \notin \dom{\ans \xi}$, then $\val tX\xi$ is
    undefined, and $q$ is the unique query such that $\xi\vdash^t_Xq$.
    \end{ls}
  \end{lsnum}
\qed\end{df}

The semantics of guards, unlike that of terms, depends not only
on the answer function but also on the preorder in the history.
Another difference from the term case is that the values of
guards, when defined, are always Boolean values.

\begin{df}[Semantics of guards]   \label{guard-sem-def}
Let $\phi$ be a guard and $\xi$ a history in an $\U$-structure
$X$.
\begin{lsnum}
  \item If $\phi$ is a Boolean term, then its value (if any) and
  causality relation are already given by
  Definition~\ref{term-val-def}.
  \item If $\phi$ is $(s\preceq t)$ and if both $s$ and $t$ have values
  with respect to $\xi$, then $\val\phi X\xi=\ttt{true}$ if, for every
  initial segment $\eta\initeq\xi$ such that \val tX\eta\ is defined,
  $\val sX\eta$ is also defined.  Otherwise, $\val\phi
  X\xi=\ttt{false}$.  Also declare that $\xi\vdash^\phi_Xq$ for no
  $q$.
  \item If $\phi$ is $(s\preceq t)$ and if $s$ has a value with respect
  to $\xi$ but $t$ does not, then define $\val\phi X\xi$ to be
  \ttt{true}; again declare that $\xi\vdash^\phi_Xq$ for no $q$.
  \item If $\phi$ is $(s\preceq t)$ and if $t$ has a value with respect
  to $\xi$ but $s$ does not, then define $\val\phi X\xi$ to be
  \ttt{false}; again declare that $\xi\vdash^\phi_Xq$ for no $q$.
  \item If $\phi$ is $(s\preceq t)$ and if neither $s$ nor $t$ has a
  value with respect to $\xi$, then $\val\phi X\xi$ is undefined,
  and $\xi\vdash^\phi_Xq$ if and only if $\xi\vdash^s_Xq$ or
  $\xi\vdash^t_Xq$.
  \item If $\phi$ is $\psi_0\kand\psi_1$ and both $\psi_i$ have value
  \ttt{true}, then $\val\phi X\xi=\ttt{true}$ and no query is
  produced.
  \item If $\phi$ is $\psi_0\kand\psi_1$ and at least one $\psi_i$ has
  value \ttt{false}, then $\val\phi X\xi=\ttt{false}$ and no query
  is produced.
  \item If $\phi$ is $\psi_0\kand\psi_1$ and one $\psi_i$ has value
  \ttt{true} while the other, $\psi_{1-i}$, has no value, then
  $\val\phi X\xi$ is undefined, and $\xi\vdash^\phi_Xq$ if and only if
  $\xi\vdash^{\psi_{1-i}}_Xq$.
  \item If $\phi$ is $\psi_0\kand\psi_1$ and neither $\psi_i$ has a
  value, then \val\phi X\xi\ is undefined, and $\xi\vdash^\phi_Xq$
  if and only if $\xi\vdash^{\psi_i}_Xq$ for some $i$.
  \item The preceding four clauses apply with $\kor$ in place of
  $\kand$ and \ttt{true} and \ttt{false} interchanged.
  \item If $\phi$ is $\neg\psi$ and $\psi$ has a value, then
  $\val\phi X\xi=\neg\val\psi X\xi$ and no query is produced.
  \item If $\phi$ is $\neg\psi$ and $\psi$ has no value then
  \val\phi X\xi\ is undefined and $\xi\vdash^\phi_Xq$ if and only
  if $\xi\vdash^{\psi}_Xq$.
\end{lsnum}
\qed\end{df}

The semantics of a rule, for an $\U$-structure $X$, an
appropriate template assignment, and a history $\xi$, consists
of a \emph{causality relation}, declarations of whether $\xi$
is \emph{final} and whether it \emph{succeeds} or \emph{fails},
and a set of \emph{updates}.

\begin{df}[Semantics of Rules] \label{rule-sem-def}
Let $R$ be a rule and $\xi$ a history
  for the $\U$-structure $X$.  In the following clauses, whenever we
  say that a history succeeds or that it fails, we implicitly also
  declare it to be final; contrapositively, when we say that a history
  is not final, we implicitly also assert that it neither succeeds nor
  fails.
  \begin{lsnum}
    \item If $R$ is an update rule $f(t_1,\dots,t_n):=t_0$ and if all
    the $t_i$ have values $\val{t_i}X\xi=a_i$, then $\xi$ succeeds for
    $R$, and it produces the update set
    $\{\sq{f,\sq{a_1,\dots,a_n},a_0}\}$ and no queries.
    \item If $R$ is an update rule $f(t_1,\dots,t_n):=t_0$ and if some
    $t_i$ has no value, then $\xi$ is not final for $R$, it produces
    the empty update set, and $\xi\vdash^R_Xq$ if and only if
    $\xi\vdash^{t_i}_Xq$ for some $i$.
  \item If $R$ is $\ttt{issue\,}f(t_1,\dots,t_n)$ and if all the $t_i$
  have values $\val{t_i}X\xi=a_i$, then $\xi$ succeeds for $R$, it
  produces the empty update set, and $\xi\vdash^R_Xq$ for the single
  query $q=\hat f[a_1,\dots,a_n]$ provided $q\notin\dom{\ans\xi}$; if
  $q\in\dom{\ans\xi}$ then no query is produced.
    \item If $R$ is $\ttt{issue\,}f(t_1,\dots,t_n)$ and if some $t_i$
    has no value, then $\xi$ is not final for $R$, it produces the
    empty update set, and $\xi\vdash^R_Xq$ if and only if
    $\xi\vdash^{t_i}_Xq$ for some $i$.
  \item If $R$ is \ttt{fail}, then $\xi$ fails for
    $R$; it produces the empty update set and no queries.
    \item If $R$ is a conditional rule $\ttt{if\ } \phi \ttt{\ then\ }
R_0 \ttt{\ else\ } R_1 \ttt{\ endif}$ and if $\phi$ has no
value, then $\xi$ is not final for $R$, and it produces the
empty update set. $\xi\vdash^R_Xq$ if and only if
$\xi\vdash^\phi_Xq$.
     \item If $R$ is a conditional rule $\ttt{if\ } \phi
\ttt{\ then\ } R_0 \ttt{\ else\ } R_1 \ttt{\ endif}$ and if
$\phi$ has value \ttt{true} (resp.\ \ttt{false}), then
finality, success, failure, updates, and queries are the same
for $R$ as for $R_0$ (resp.\ $R_1$).
     \item If $R$ is a parallel combination $\ttt{do\ in\ parallel\
     }R_1,\dots,R_k\ttt{\ enddo}$ then:
       \begin{ls}
     \item $\xi\vdash^R_Xq$ if and only if $\xi\vdash^{R_i}_Xq$
     for some $i$.
          \item The update set for $R$ is the union of the update sets
     for all the components $R_i$.  If this set contains two
     distinct updates at the same location, then we say that a
     \emph {clash} occurs (for $R$, $X$, and $\xi$).
     \item $\xi$ is final for $R$ if and only if it is final for
     all the $R_i$.
     \item $\xi$ succeeds for $R$ if and only if it succeeds for
         all the $R_i$ and no clash occurs.
     \item $\xi$ fails for $R$ if and only if it is final for $R$
     and either it fails for some $R_i$ or a clash occurs.
       \end{ls}
  \end{lsnum}
\qed\end{df}

\begin{df} \label{asm-next_state}
Fix a rule $R$ endowed with a template assignment, and let $X$
be an $\U$-structure and $\xi$ be a history for $X$.  If $\xi$
is successful and final for $R$ over $X$, then the
\emph{successor} $\tau(X,\xi)$ of $X$ with respect to $R$ and
$\xi$ is defined from the update set $\DD(X,\xi)$ as in the
Step Postulate, Part~C.
\end{df}

It is easy to check (see \cite[Part~II, Lemma~3.18]{general})
that $\tau$ is well-defined; $\Delta^+$ will not prescribe two
contradictory updates of the same location under a successful,
final history.

\begin{df}   \label{asm-def}
  An \emph{interactive, small-step, ASM} consists of
  \begin{ls}
    \item an ASM program $\Pi$ in some vocabulary $\U$,
    \item a nonempty set $\scr S$ of $\U$-structures called \emph{states}
    of the ASM, and
    \item a nonempty set $\scr I\subseteq \scr S$ of \emph{initial states},
  \end{ls}
subject to the requirements that \scr S and \scr I are closed
under isomorphism and that \scr S is closed under transitions
in the following sense.  If $X\in\scr S$ and if $\xi$ is a
successful, final history for $\Pi$ in $X$, then the successor
$\tau(X,\xi)$ of $X$ with respect to $\Pi$ and $\xi$ is also in
\scr S.    \qed
\end{df}

It is shown in \cite[Part~II]{general} that ASMs are
algorithms, in the sense defined above by the postulates, and
that, conversely, all algorithms are behaviorally equivalent to
ASMs.

\section{Impatience}    \label{impat}

We call an algorithm \emph{patient} if it never finishes a step
until the environment has answered all queries from that step.
(It patiently waits for answers to all its queries.) Formally,
this means that all final, attainable histories are complete.
Otherwise, we call the algorithm \emph{impatient}.

A query is said to be \emph{blocking} if, once it is issued,
the algorithm's step cannot end without a reply to this query.
Thus, an algorithm is patient if and only if all its queries
are blocking.

In this paper, we are concerned with non-blocking queries, and
specifically with the possibility that the reply to such a
query may arrive and be used by the algorithm in a later step
than the one that produced the query.  The present section
describes where, in an ASM program, non-blocking queries can
originate.  Of course, we must first say precisely what it
means for a query to originate in a particular part of a rule
--- or of a term, or of a guard.

We present the material in this section in the context of the
traditional ASM syntax and semantics described above.  It applies
equally, however, to the modified syntax and semantics that was
described in Section~\ref{soln} and will be formalized in
Section~\ref{locate}.  The changes we introduce do not affect the
proofs in the present section.

The discussion will be
simplified by the following definitions and convention.

\begin{df}
Let $S$ be a term or a guard or a rule, let $X$ be a state, and
let $\xi$ be a history for $X$.  We define
\begin{align*}
\Issued^S_X(\xi)&=\{q:(\exists\eta\initeq\xi)\,\eta\vdash^S_Xq\}\\
\text{Pending}^S_X(\xi)&=\Issued^S_X(\xi)-\dom{\ans\xi}.\qed
\end{align*}
\end{df}

Note that, in the case of a rule, this definition agrees with
Definition~\ref{iss-pend} for algorithms.  We are just
extending the ``Issued'' and ``Pending'' notation to apply also
to terms and guards (and adding the superscript $S$, which was
unnecessary earlier because the role of $S$ was played there by
a fixed algorithm). The next definition also extends to terms
and guards terminology already available for rules.

\begin{df}
Let $S$ be a term or a guard, let $X$ be a state, and let $\xi$
be a history for $X$.  We say that the history $\xi$ is
\emph{final} for $S$ in $X$ if \val SX\xi\ is defined.
\end{df}

\begin{conv}
When we speak of syntactic parts of a term, guard, or rule, we
mean occurrences of those syntactic parts.  Thus, for example,
``subrule'' really means ``occurrence of subrule.''
\end{conv}

We now define the origins of a query caused by a term or guard
or rule.  The definition involves going systematically through
the definitions of the semantics of term, guards, and rules
(Definitions~\ref{term-val-def}, \ref{guard-sem-def}, and
\ref{rule-sem-def}), and checking all the clauses where a query
is caused. For the reader's convenience, we append to some
clauses of the following definition some additional
information, in brackets, about the circumstances in which
those clauses can apply.  These bracketed comments can easily
be verified by inspection of Definitions~\ref{term-val-def},
\ref{guard-sem-def}, and \ref{rule-sem-def}.

\begin{df}  \label{origin}
Let $X$ be a state, and $\xi$ a history for it, and $q$ a
potential query.
\begin{ls}
\item If $t$ is a term $f(t_1,\dots,t_n)$, if \val{t_i}X\xi\ is
    undefined for at least one $i$, and if $\xi\vdash_X^tq$,
    then the origins of $q$ in $t$ are the origins of $q$ in
    all those $t_i$ for which $\xi\vdash_X^{t_i}q$.
\item If $t$ is a term $f(t_1,\dots,t_n)$, if \val{t_i}X\xi\ is
    defined for all $i$, and if $\xi\vdash_X^tq$, then $q$ has
    exactly one origin in $t$, namely $t$ itself.  [Here $f$ is an
    external function symbol and $q$ is the q-value of $t$.]
\item If $\phi$ is $(s\preceq t)$ and $\xi\vdash^\phi_Xq$, then
    the origins of $q$ in $\phi$ are its origins in $s$ (if any,
    i.e., if $\xi\vdash^s_Xq$) and its origins in $t$ (if
    any).  [According to the semantics of guards, if either $s$ or $t$
    has a value, then $(s\preceq t)$ issues no queries.  So the
    present clause applies only when $\xi$ is not final for either of
    these terms.]
\item If $\phi$ is $\psi_0\kand\psi_1$ or $\psi_0\kor\psi_1$
    and $\xi\vdash^\phi_Xq$, then the origins of $q$ in $\phi$
    are its origins in $\psi_0$ (if any) and its origins in
    $\psi_1$ (if any).  [At most one of $\psi_0$ and $\psi_1$
    has a value under $\xi$,
    and if one does then that value is
    \ttt{true} in the case of $\kand$ and \ttt{false} in the case of
    $\kor$.]
\item If $\phi$ is $\neg\psi$ and $\xi\vdash^\phi_Xq$, then the
    origins of $q$ in $\phi$ are the same as in $\psi$.
\item If $R$ is an update rule $f(t_1,\dots,t_n):=t_0$ and
    $\xi\vdash^R_Xq$, then the origins of $q$ in $R$ are the
    origins of $q$ in all those $t_i$ for which
    $\xi\vdash^{t_i}_Xq$.
\item If $R$ is $\ttt{issue\,}f(t_1,\dots,t_n)$, if all the
    $t_i$ have values $\val{t_i}X\xi=a_i$, and if $\xi\vdash^R_Xq$,
    then $q$ has
    exactly one origin in $R$, namely $f(t_1,\dots,t_n)$.  [Here $f$
    is an external function symbol and $q$ is the q-value of
    $f(t_1,\dots,t_n)$.]
\item If $R$ is $\ttt{issue\,}f(t_1,\dots,t_n)$, if
    some $t_i$ has no value, and if $\xi\vdash^R_Xq$, then
    the origins of $q$ in $R$ are the
    origins of $q$ in all those $t_i$ for which
    $\xi\vdash^{t_i}_Xq$.
\item If $R$ is a conditional rule $\ttt{if\ } \phi \ttt{\
    then\ } R_0 \ttt{\ else\ } R_1 \ttt{\ endif}$, if $\phi$
    has no value under $\xi$,
    and if $\xi\vdash^R_Xq$, then
    the origins of $q$ in $R$ are the
    origins of $q$ in $\phi$.
\item If $R$ is a conditional rule $\ttt{if\ } \phi \ttt{\
    then\ } R_0 \ttt{\ else\ } R_1 \ttt{\ endif}$, if $\phi$
    has value \ttt{true} (resp.\ \ttt{false}), and if
    $\xi\vdash^R_Xq$, then the origins of $q$ in $R$ are its
    origins in $R_0$ (resp.\ $R_1$).
\item If $R$ is a parallel combination $\ttt{do\ in\ parallel\
     }R_1,\dots,R_k\ttt{\ enddo}$ and if
    $\xi\vdash^R_Xq$, then the origins of $q$ in $R$ are its
    origins in all those $R_i$ for which $\xi\vdash^{R_i}_Xq$.
\end{ls}
\end{df}

In the preceding definition, $X$ and $\xi$ were fixed and were
therefore not mentioned in the ``origin'' terminology.
\looseness=-1 When necessary, we make them explicit by a phrase
like ``origin of $q$ in $R$ with respect to $X$ and $\xi$.''

\begin{la}  \label{orig}
Let $S$ be a term or guard or rule, let $X$ be a state, let
$\xi$ be a history for $X$, and let $q$ be a potential query in
$X$.  Then $\xi\vdash^S_Xq$ if and only if $q$ has at least one
origin in $S$ with respect to $X$ and $\xi$.  Any origin of $q$
is a query-term $t$, a subterm of $S$, with $\qval tX\xi=q$.
Furthermore, this $t$ is also the (unique) origin of $q$ in
$t$; in particular, $\xi\vdash^t_Xq$.
\end{la}

\begin{pf}
Proceed by induction, first on terms, then on guards, and
finally on rules.  In every case, the proof is just a
comparison of the definition of ``origin'' with the parts of
Definitions~\ref{term-val-def}, \ref{guard-sem-def}, and
\ref{rule-sem-def} that describe the causality relation.
\end{pf}

After these preliminaries, we can look in detail at impatience,
the phenomenon of an ASM's step ending even though some of its
queries have not been answered.  In more detail, the phenomenon
involves five entities:
\begin{ls}
\item an ASM program $\Pi$,
\item a state $X$ of $\Pi$,
\item a history $\xi$ that is final for $X$ with respect to
    $\Pi$ (so the step ends),
\item a query $q\in\text{Pending}^\Pi_X(\xi)$ (so $q$ has been
    issued but not answered during this step), and
\item an initial segment $\eta\initeq\xi$ such that
    $\eta\vdash^\Pi_Xq$.
\end{ls}
In connection with the last of these items, $\eta$, recall that
for $q$ to be issued during a step where the history is $\xi$
it must be caused by some initial segment of $\xi$, though not
necessarily by $\xi$ itself.

We have simplified the notation by using the same symbol $\Pi$
for an ASM program and for its underlying rule, even though the
program also includes additional materials, particularly the
template assignment. This additional material will remain
fixed, so our abuse of notation will not cause confusion.

The following proposition describes the possible origins of
queries that remain unanswered at the end of a step.  Notice
that, when the $S$ in the proposition is a rule $\Pi$, then the
hypotheses of the proposition describe the five items listed above.

\begin{prop}    \label{source}
Let $S$ be a term or guard or rule.  Let $X$ be a state and let
$\eta\initeq\xi$ be two histories for $X$, such that $\xi$ is
final for $X$ with respect to $S$.  Let $q$ be a query such
that $\eta\vdash^S_Xq$ but $q\notin\dom{\dot\xi}$.  Then all
origins of $q$ in $S$ with respect to $X$ and $\eta$ are of one
of the following sorts:
\begin{ls}
\item query-subterms of $s$ or $t$ in a timing guard $s\preceq
    t$
    within $S$,
\item query-subterms of $\psi_0$ or $\psi_1$ in a
    Kleene-conjunction $\psi_0\kand\psi_1$ or
    Kleene-disjunction $\psi_0\kor\psi_1$ within $S$,
\item arguments $t$ of issue-rules $\ttt{issue\,}(t)$ within
    $S$.
\end{ls}
\end{prop}

\begin{pf}
Assume that $S,X,\eta,\xi$, and $q$ are as in the hypothesis of
the proposition and that $o$ is an origin of $q$ in $S$ with
respect to $X$ and $\eta$.  Assume also, as an induction
hypothesis, that the proposition becomes true if $S$ is
replaced by any proper subterm, subguard, or subrule (while
$X,\eta,\xi,q$, and $o$ are unchanged.)

To save a little writing later, observe that the hypothesis
that $\eta\vdash^S_Xq$ is redundant, because, according to
Lemma~\ref{orig}, if it didn't hold then there would be no
origin of $q$ in $S$ with respect to $X$ and $\eta$, and so the
conclusion of the proposition would hold vacuously.

The proof will repeatedly use the observation that, if the
conclusion of the proposition holds when $S$ is replaced by
some subterm, subguard, or subrule $S'$ of $S$, then it also
holds for $S$ itself.  The reason is that the conclusion refers
to $S$ only in the context of saying that some guard or rule
occurs within $S$.  If we find the desired guard or rule within
$S'$ then we certainly have it within $S$.

The fact that $\eta\vdash^S_Xq$ must arise from one of the
clauses of Definition~\ref{term-val-def}, \ref{guard-sem-def},
or \ref{rule-sem-def}, with $\eta$ in place of the $\xi$ in the
definition.  And this clause cannot be one of the many clauses
where the definition says that no query is caused, i.e.,
clause~2 and the first case in clause~3 of
Definition~\ref{term-val-def}; clauses~2, 3, 4, 6, 7, and 11 as
well as the part of clause~10 analogous to clause~7 in
Definition~\ref{guard-sem-def}; and clauses~1 and 5 of
Definition~\ref{rule-sem-def}.  We examine the remaining
possibilities in turn, labeling them according to the clause in
Definition~\ref{term-val-def}, \ref{guard-sem-def}, or
\ref{rule-sem-def} that provides $\eta\vdash^S_Xq$.

\smallskip

\textbf{\ref{term-val-def}-1}: $S$ is a term $f(t_1,\dots,t_n)$
and, for at least one $i$, \val{t_i}X\eta\ is undefined.
According to Definition~\ref{origin}, $o$ is an origin of $q$
in some $t_i$ (with respect to $X$ and $\eta$).  For such an
$i$, $\xi$ will be final with respect to $t_i$, i.e.,
\val{t_i}X\xi\ will be defined, because otherwise, the same
clause of Definition~\ref{term-val-def} (now applied to $\xi$
rather than $\eta$) would contradict the assumption that $\xi$
is final for $S$.  Thus, the hypotheses of the proposition are
satisfied with $t_i$ in place of $S$.  By induction hypothesis,
the conclusions of the proposition hold for $t_i$, and, as
observed above, it immediately follows that they also hold for
$S$.  (Though it isn't needed for the proof, it may help the
reader if we point out that this case cannot actually occur.
Indeed, by what we have just proved, we would have a guard
(involving $\prec$ or $\kand$ or $\kor$) or an issue-rule
within a term, and this cannot happen in the ASM syntax.)

\smallskip

\textbf{\ref{term-val-def}-3, second part}: $S$ is a term
$f(t_1,\dots,t_n)$ where $f$ is an external function symbol;
each $t_i$ has a value $\val{t_i}X\eta=a_i$; and
\[
q=\qval SX\eta=\hat f[a_1,\dots,a_n].
\]
(This clause in Definition~\ref{term-val-def} also says that
$q\notin\dom{\dot\eta}$, but this is immediate from the
assumptions that $q\notin\dom{\dot\xi}$ and $\eta\initeq\xi$.)
Lemma~3.6 of \cite[Part~II]{general} gives us that, when we
pass from $\eta$ to its extension $\xi$, the values of the
$t_i$'s do not change.  So we have $\val{t_i}X\xi=a_i$ and
therefore (by the same clause of Definition~\ref{term-val-def})
\[
\qval SX\xi=\hat f[a_1,\dots,a_n]=q.
\]
But then, since $q\notin\dom{\dot\xi}$, the same clause tells
us that $\xi$ is not final for $S$.  This contradicts the
hypothesis of the proposition, so this case simply cannot
arise.

\smallskip

\textbf{\ref{guard-sem-def}-1}: Here $S$ is guard that is a
Boolean term, so this case is included in the cases already
treated where $S$ is a term.

\smallskip

\textbf{\ref{guard-sem-def}-5}: Here $S$ is a timing guard
$(s\preceq t)$ (and neither of the terms $s,t$ has a value with
respect to $\eta$).  By Lemma~\ref{orig}, $o$ is a subterm of
this timing guard, and so we have the first of the three
alternatives in the conclusion of the proposition.

\smallskip

\textbf{\ref{guard-sem-def}-8 or 9}:  Here $S$ is a Kleene
conjunction, and so its subterm $o$ satisfies the second
alternative in the proposition.

\smallskip

\textbf{\ref{guard-sem-def}-10}:  Here $S$ is a Kleene
disjunction, and so we again get the second alternative of the
proposition.

\smallskip

\textbf{\ref{guard-sem-def}-12}: Here $S$ is $\neg\psi$.  By
definition, origins in $S$ are the same as in $\psi$.  Also, by
definition, since $\xi$ isn't final for $S$, it isn't final for
$\psi$.  Thus, the induction hypothesis applies and tells us
that the conclusion of the proposition holds with $\psi$ in
place of $S$.  But then it also holds for $S$.

\smallskip

\textbf{\ref{rule-sem-def}-2}: $S$ is an update rule
$f(t_1,\dots,t_n):=t_0$ and not all $t_i$ have values with
respect to $\eta$.  The argument here is essentially the same
as for case~\ref{term-val-def}-1 above. $o$ is an origin of $q$
in some $t_i$, and $\xi$ must be final for $t_i$ as otherwise
it would not be final for $S$.  By induction hypothesis, the
conclusion of the proposition holds with $t_i$ in place of $S$,
and therefore it also holds for $S$.

\smallskip

\textbf{\ref{rule-sem-def}-3}: Here $S$ is an issue-rule and,
by definition of ``origin,'' $o$ is its argument.  So we have
the third alternative in the proposition.

\smallskip

\textbf{\ref{rule-sem-def}-4}: The argument here is again
essentially the same as for cases~\ref{term-val-def}-1 and
\ref{rule-sem-def}-2; we spare the reader (and ourselves) a
third occurrence of this same argument.

\smallskip

\textbf{\ref{rule-sem-def}-6}: Here $S$ is a conditional rule
whose guard $\phi$ has no value with respect to $\eta$.  By
definition, $o$ is an origin of $q$ in $\phi$ with respect to
$\eta$.  Furthermore, $\xi$ must be final for $\phi$, because
otherwise it could not be final for $S$.  So the induction
hypothesis applies and we get the conclusion of the proposition
with $\phi$ in place of $S$, and therefore also for $S$.

\smallskip

\textbf{\ref{rule-sem-def}-7}: $S$ is a conditional rule
$\ite\phi{R_0}{R_1}$ and $\phi$ has a value with respect to
$\eta$. We assume $\val\phi X\eta=\ttt{true}$; the case of
\ttt{false} is the same with $R_0$ and $R_1$ interchanged.  By
definition, the origins of $q$ in $S$ are the same as in $R_0$.
Also, by Lemma~3.12 of \cite[Part~II]{general}, $\val\phi
X\xi=\ttt{true}$, so the finality of $\xi$ for $S$ implies that
$\xi$ is also final for $R_0$.  So the induction hypothesis
applies and gives us the conclusion of the proposition with
$R_0$ in place of $S$.  As usual, the conclusion for $S$
follows.

\smallskip

\textbf{\ref{rule-sem-def}-8}: $S$ is a parallel combination
with components $R_i$.  The definition of ``origin'' says that
$o$ is an origin of $q$ in at least one of the $R_i$.  And
$\xi$ must be final for that $R_i$ because otherwise it could
not be final for $S$.  So the induction hypothesis gives us the
conclusion of the proposition with $R_i$ in place of $S$, and
the conclusion for $S$ follows.
\end{pf}

\begin{rmk}
In view of Proposition~\ref{source}, we can limit the use of
the new syntax $\ql{g(\bld u)}{f(\bld t)}$ to the places
described in the proposition, namely subterms of timing guards,
of Kleene conjunctions, and of Kleene disjunctions, and arguments
of issue-rules.  External function symbols occurring anywhere
else in an ASM program produce blocking queries, so there is no
need to provide locations for late replies.  And if a
reply-location is provided for a blocking query, with the
intention of having an on-time reply recorded there, then the
program can easily be altered so that the ASM reads the reply
in its history and writes it into the desired location.
\end{rmk}

\begin{ex}  \label{source3}
Here are some trivial examples showing that all the
alternatives in the conclusion of Proposition~\ref{source} can
occur (with $S$ being a rule).  Assume that the vocabulary has
three external, nullary function symbols $a,b,c$ and that the
templates $\hat a,\hat b,\hat c$ assigned to them are distinct.

For the first alternative in Proposition~\ref{source}, consider
the rule
\[
\ite{a\prec b}{x:=1}{x:=2}.
\]
The empty history causes both $\hat a$ and $\hat b$. Any history $\xi$
with domain $\{\hat a\}$ is final and has $\hat b$ pending.  This
$\hat b$ has exactly one origin, namely the unique $b$ in the
rule. (The updates of $x$ could be replaced by certain other rules,
for example \ttt{skip}, without affecting the idea.)  The same program
also serves as an example if the reply for $b$ arrives before that for
$a$; then the history with only the reply for $b$ is final, $a$ is
pending, and its only origin is in the timing guard $a\prec b$.

For the second alternative, consider
\[
\ite{(a=b)\kand(a=c)}{x:=1}{x:=2}.
\]
The empty history causes all three of $\hat a,\hat b,\hat c$.
Any history $\xi$ with domain $\{\hat a,\hat b\}$ and with
$\xi(\hat a)\neq\xi(\hat b)$ is final and has $\hat c$ pending.
The only origin of $\hat c$ in this rule is the unique
occurrence of $c$.

There is an analogous example with $\kor$ in place of $\kand$.
Just use a $\xi$ that gives the same reply to the two queries
$\hat a$ and $\hat b$.

Finally, for the third alternative, just use the rule
$\ttt{issue }(a)$.  The empty history causes $\hat a$ and is
final, with $\hat a$ pending.
\end{ex}

\begin{rmk}
We take this opportunity to clarify Remark~3.17 of
\cite[Part~II]{general}, which begins: ``Issue rules are the
only way an ASM can issue a query without necessarily waiting
for an answer.''  This appears to deny the possibility of the
first two alternatives in Proposition~\ref{source}.  Indeed, if
``waiting for an answer'' means ``waiting until an answer is
received,'' then the examples just given show that this is
wrong. It becomes correct, however, if ``waiting for an
answer'' means ``waiting at least for a moment,'' i.e., not
finishing the step immediately.  Note that, in the parts of
Example~\ref{source3} that don't use \ttt{issue}, the pending
query is caused not by the final history but by a proper
initial segment.  (In the notation of Proposition~\ref{source},
$\eta\neq\xi$.)  In this sense, the ASM does wait after issuing
the query and before finishing the step.

Another description of what happens in these examples is that
the unanswered query $q$ is issued by the final history $\xi$
(in the sense of being in $\text{Issued}(\xi)$) but it is not
caused by the final history ($\xi\not\vdash q$).  In the second
sentence of Remark~3.17, we used the phrase ``a history causes
a rule to issue a query,'' which is ambiguous in view of the
difference between causing and issuing.  It should be interpreted as
causing, not merely issuing.
\end{rmk}

\section{Announcing Locations for Late Replies}   \label{locate}

In this section, we present the small modifications of
\cite{general} needed to accommodate the \sq{q,\rl,l} method,
proposed in Section~\ref{soln}, for handling persistent queries
in ASMs.  Some of these modifications directly affect the
syntax and semantics of ASMs or (in one case) even the notion
of algorithm from \cite{general}; we exhibit these with the
heading ``Modification.''  If these are violated, then our ASM
programs with persistent queries won't make sense.  Other
modifications describe what we expect to see in programs and in
the environment's behavior.  These concern either constraints
on the environment or good programming practice; we label these
``Intention.''  If they are violated, ASM programs with
persistent queries will still make sense, but it may not be the
sense that was intended.

\begin{mdf} \label{mdf-rl}
The set $\Lambda$ of labels contains the ``reply location
marker'' \rl.
\end{mdf}

The purpose of this modification is of course to ensure
availability of the queries \sq{q,\rl,l} that we want to use
when issuing a persistent query $q$ with reply location $l$. It
may seem that we should also require that all dynamic function
symbols $f$ should be among the labels, so that they can be
used in the location part $l$ of \sq{q,\rl,l}.  This
requirement would do no harm, but it may be overkill, since
there may be many dynamic function symbols that will not be
used for reply locations in a particular program.  Accordingly,
we do not impose this requirement but instead use the following
definition to keep track of which function symbols are
available to serve as the first component of a reply location.

\begin{df}
A function symbol is \emph{reply-available} if it is dynamic
and is also a member of the set $\Lambda$ of labels.
\end{df}

\begin{rmk}
We have insisted here that the function-symbol component of a
reply location be dynamic.  This is not strictly necessary; one
could imagine using a static function --- one that the
algorithm can never update --- in this role, since the updates
would be done by the environment, not by the algorithm.  But it
seems strange to allow this when the update is being done at
the request of the algorithm.

Notice, for example, the following undesirable consequence of
allowing reply locations that begin with a static function.
Suppose the environment can provide echoes; that is, an
algorithm can issue a query of the form ``answer this with
$x$,'' where $x$ is an element of the state, and get reply $x$.
Then by issuing the query
\[
\text{answer this with }x\,[=:f(\bld t)],
\]
the algorithm can achieve (after the end of the current step)
the effect of the update $f(\bld t):=x$.  That should not be
possible when $f$ is static.
\end{rmk}

\begin{rmk}
The definition of reply-available is designed to cohere with
our convention in Section~\ref{soln} about the format of the
queries that provide reply locations.  Had we chosen a
different format, for example using some codes for the function
symbols, then the definition should be modified accordingly.
\end{rmk}

Our next task is to understand, in a way that fits the general
notions of algorithms and ASMs, the external function calls
accompanied by reply locations.  We can fit this syntactic
construct
\[
\ql{g(u_1,\dots,u_m)}{f(t_1,\dots,t_n)}
\]
into the ASM framework by treating it as a new external
function symbol with all of $u_1,\dots,u_m,t_1,\dots,t_n$ as
arguments.  That is, we require the availability of a new
$(m+n)$-ary function symbol, which we denote by $\ql gf$, and
we treat $\ql{g(\bld u)}{f(\bld t)}$ as syntactic sugar for
$\ql gf(\bld u,\bld t)$.  The following modification and
definition formalize this convention.

\begin{mdf}
For certain pairs $g,f$, where $g$ is an external function
symbol and $f$ a reply-available function symbol, an external
function symbol $\ql gf$ is designated, with arity equal to the
sum of the arities of $g$ and $f$.
\end{mdf}

\begin{df}
When $\ql gf$ is defined, we say that $f$ is
\emph{reply-available for} $g$.  In this case, if $g$ is
$m$-ary and $f$ is $n$-ary, then
\ql{g(u_1,\dots,u_m)}{f(t_1,\dots,t_n)} means $\ql
gf(u_1,\dots,u_m,t_1,\dots,t_n)$.
\end{df}

At this stage, we have ensured that ASM programs written with
the \ql{g(\bld u)}{f(\bld t)} notation are syntactically
correct, provided $f$ is reply-available for $g$.  As a first
step toward semantic correctness, we want them to issue the
right queries.

\begin{intent}
When \ql gf is defined, the associated template is
\[
\widehat{\ql gf}=\sq{\hat g,\rl,f,\#(m+1),\dots,\#(m+n)},
\]
where $g$ is $m$-ary and $f$ is $n$-ary.
\end{intent}

\begin{rmk}
We have, once again, taken some liberties with the bracketing.
Without liberties, we would have $\hat
g^\frown\sq{\rl,f,\#(m+1),\dots,\#(m+n)}$, where ${}^\frown$
denotes concatenation of sequences.  It may also be worth
noting that $\hat g$ is the same as $\hat g[\#1,\dots,\#m]$.
\end{rmk}

The next modification says that, when the algorithm issues a
query that contains the \rl\ label, it should get an answer to
the ``query part'' preceding \rl, since the rest merely
specifies a reply location.  It turns out that the only change
needed in the definitions and postulates from \cite{general} is
that, when \sq{q,\rl,l} is caused by an initial segment of a
history $\xi$, it is not this query itself but rather the
initial segment $q$ that counts as issued.

\begin{mdf}
The definition of $\Issued_X(\xi)$ in Definition~\ref{iss-pend}
is amended as follows.  $\Issued_X(\xi)$ consists of those
queries $q$ such that
\begin{ls}
\item $q$ does not contain \rl, and
\item for some initial segment $\eta$ of $\xi$, either
    $\eta\vdash_Xq$ or $\eta\vdash_X\sq{q,\rl,l}$ for some
    sequence $l$.
\end{ls}
\end{mdf}

\begin{intent}
The only external function symbols whose templates contain \rl\
are those of the form \ql gf.
\end{intent}

The preceding ``Intentions'' imply that no template contains
more then one occurrence of \rl.  Nevertheless, our
modification of the definition of Issued can handle queries
with several \rl's; the first occurrence of \rl\ is the one
that counts.

\begin{rmk}
It is possible for an ASM program to prescribe two different
reply locations for what turns out to be the same query.  For
example, we might have both \ql{g(\bld u)}{f(\bld t)} and
\ql{g(\bld u')}{f'(\bld t')} where, in some (or even every)
state $\bld u$ and $\bld u'$ have the same values \bld a but
$f\neq f'$. Then the queries resulting from these two
occurrences are different, but they differ only after the \rl.
So our redefinition of Issued says that only a single query is
issued, namely $\hat g[\bld a]$.  According to Intention~\ref{replies}
below, a reply to this single query is to be written into both of the
reply locations.

To see that this is as it should be, consider the ASM program
(in the traditional sense) that results from deleting all the
reply locations.  There, $g(\bld u)$ and $g(\bld u')$ would
issue only a single query, $\hat g[\bld a]$.  Our modifications
and definitions ensure that the ASM with reply locations
behaves, in this respect, the same as the one without reply
locations.
\end{rmk}

It remains only to formally state, as intentions, the
constraints that the environment should obey in order to make
our ASMs with persistent queries behave as intended.

\begin{intent}   \label{replies}
If the algorithm has produced the query \sq{q,\rl,l}, thereby
issuing $q$, and if $l$ is a location, then the answer to $q$
should be written into location $l$.
In case
several answers are to be put into the same $l$ at the same
time, the environment chooses one of them arbitrarily.  The
environment should not write into reply locations except as
prescribed here.
\end{intent}

Usually, a program will not use the same reply location for
several different queries, and so the need for an arbitrary
choice will not arise.  If, however, the program does assign
the same reply location to several queries, then not only might
it encounter the nondeterminism described here, but replies 
might be overwritten.  

Note that, in telling the algorithm to write replies into the
prescribed locations, we have made no exception for on-time
replies.  If a query is answered during the same step in which
it was issued, then the reply goes into both the history of
that step and (when the step ends) the reply location.

\begin{intent}
Replies to persistent queries are different from \ttt{undef}.
\end{intent}

The point of this is to enable an algorithm to detect whether a
query has received a late reply.  If the value of the reply
location is initialized to \ttt{undef} and is not updated
otherwise than by a reply to $q$, then the presence of a reply
can be detected by comparing the value of this location to
\ttt{undef}.

\begin{rmk}
We briefly indicate an alternative approach that does not
require the environment to reinterpret \sq{q,\rl,l} as the
query $q$ (and does not require us to redefine Issued).  In
this approach, \ql{g(\bld u)}{f(\bld t)} should produce two
queries, namely the query $q$ that would be issued by $g(\bld
u)$ alone and the additional query \sq{q,\rl,l} giving its
reply location $l$.  The environment treats $q$ like any other
query, answering it (if possible) in the usual way.  It treats
\sq{q,\rl,l} as a message, answering it with an automatic,
immediate ``OK,'' but remembering it so that it knows where to
write a reply to $q$ later.

This approach requires a modification to \cite{general} to
allow two queries to be caused at a single point in an ASM
program.  The template assignment should now be multivalued,
assigning to \ql gf both the template \sq{\hat
g,\rl,f,\#(m+1),\dots,\#(m+n)} used above and the template
$\hat g$.  A secondary modification is to allow an $m$-ary
template $\hat g$ to be used for an $(m+n)$-ary function symbol
\ql gf.
\end{rmk}

\end{document}